\documentclass[aip,pof,longbibliography,twocolumn,reprint]{revtex4-1}
\usepackage[utf8]{inputenc}
\usepackage{amssymb,amsmath}

\newcommand\beq{\begin{equation}}
\newcommand\eeq{\end{equation}}
\newcommand\beqa{\begin{eqnarray}}
\newcommand\eeqa{\end{eqnarray}}

\newcommand{\al}{\alpha}

\usepackage{xcolor}
\definecolor{darkgreen}{rgb}{0,0.6,0.0}

\newcommand{\vicente}[1]{{ #1}}
\newcommand{\RB}[1]{{ #1}}
\newcommand{\RS}[1]{{ #1}}
\newcommand{\RG}[1]{{ #1}}
\usepackage{cancel}
\usepackage[normalem]{ulem}

\usepackage{graphicx}
\begin{document}
\title{Diffusion of impurities in a moderately dense confined granular gas}

\author{Rub\'en G\'omez Gonz\'alez}
\email{ruben@unex.es}
\affiliation{Departamento de
Did\'actica de las Ciencias Experimentales y las Matem\'aticas, Universidad de Extremadura, E-10004 C\'aceres, Spain}

\author{Vicente Garz\'o}
\email{vicenteg@unex.es; \url{https://fisteor.cms.unex.es/investigadores/vicente-garzo-puertos/}}
\affiliation{Departamento de
F\'{\i}sica and Instituto de Computaci\'on Cient\'{\i}fica Avanzada (ICCAEx), Universidad de Extremadura, E-06071 Badajoz, Spain}

\author{Ricardo Brito}
\email{brito@ucm.es}
\affiliation{Departamento de Estructura de la Materia, F\'{\i}sica T\'ermica y Electr\'onica and GISC, Universidad Complutense de Madrid, E-28040 Madrid, Spain}

\author{Rodrigo Soto}
\email{rsoto@uchile.cl}
\affiliation{Departamento de
F\'{\i}sica, Facultad de Ciencias F\'{\i}sicas y Matem\'aticas, Universidad de Chile, 8370449 Santiago, Chile}

\begin{abstract}

Mass transport of impurities immersed in a
confined quasi-two-dimensional moderately dense granular gas of inelastic hard spheres is studied. The effect of the confinement on granular particles is modeled through a collisional model (the so-called $\Delta$-model) that includes an effective mechanism to transfer the kinetic energy injected by vibration in the vertical direction to the horizontal degrees of freedom of grains. The impurity can differ in mass, diameter, inelasticity, or the energy injection at collisions, compared to the gas particles. The Enskog--Lorentz kinetic equation for the impurities is solved via the Chapman--Enskog method to first order in spatial gradients for states close to the homogeneous steady state. As usual, the three diffusion transport coefficients for tracer particles in a mixture are given in terms of the solutions of a set of coupled linear integral equations which  are solved by considering the lowest Sonine approximation. The theoretical predictions for the tracer diffusion coefficient (relating the mass flux with the gradient of the number density of tracer particles) are compared with both direct simulation Monte Carlo and molecular dynamics simulations. The agreement is in general good, except for strong inelasticity and/or large contrast of energy injection at tracer-gas collisions compared to gas-gas collisions. Finally, as an application of our results, the segregation problem induced by both a thermal gradient and gravity is exhaustively analyzed.      

\end{abstract}



\date{\today}
\maketitle

\section{Introduction}
\label{sec1}

The study of transport properties of granular gases (modeled as a gas of hard spheres undergoing inelastic collisions) confined in a given direction is still a challenging open problem. In particular, one geometry that has attracted the attention of many researchers is the so-called quasi-two-dimensional geometry, namely, when the granular gas is confined in a box whose vertical $z$-direction is slightly larger than one particle diameter.\cite{PhysRevLett.81.4369,PhysRevE.70.050301,Melby_2005} When the box is subjected to vertical vibrations, energy is injected into the vertical degrees of freedom of the particles as they collide with the plates of the system. This kinetic energy gained by the particles is then transferred to the particles' horizontal degrees of freedom when they collide with each other. Under certain conditions, it has been observed that the system maintains a homogeneous fluidized state (as viewed from above) over a wide range of parameters.\cite{MS16}

However, a kinetic theory description of the above confined system is quite intricate due basically to the restrictions imposed by the confinement in the corresponding Boltzmann collisional operator. Thus, although several attempts have been recently made by considering the above approach, \cite{MBGM22,MGB22,MPGM23} one can consider in a more effective way the effect of the confinement on the dynamics of granular particles via the collisional model proposed in Ref.\ \onlinecite{BRS13}. The idea behind this collisional model is that the magnitude of the normal component of the relative velocity of the colliding particles is increased by a constant factor $\Delta$ in each collision. The term associated with the factor $\Delta$ in the corresponding collisional rule tries to mimic the transfer of kinetic energy from the vertical degrees of freedom of granular particles to the horizontal ones. This sort of collisional model will be referred to here as the $\Delta$-model.

The $\Delta$-model has been employed extensively in recent years to study monocomponent granular gases. It has been applied to analyze the homogeneous state, \cite{BGMB13,BMGB14} to obtain the Navier–Stokes transport coefficients, \cite{BBMG15} and to perform stability analyses of the time-dependent homogeneous state. \cite{BBGM16} Independently, the shear viscosity coefficient of a dilute granular gas has been explicitly calculated and shows good agreement with computer simulations. \cite{SRB14} Extensions of these works to moderate densities, within the framework of the Enskog kinetic equation, have also been reported in several papers. \cite{GBS18,GBS20,GBS21a} In addition, the $\Delta$-model has been employed in the study of systems with long-range interactions, \cite{joyce2016attractor} absorbing states, \cite{maire2024interplay} and the formation of quasi-long-range ordered phases. \cite{plati2024quasi,10.1063/5.0217958}

More recently, the $\Delta$-model has been also used to determine the Navier--Stokes transport coefficients of \emph{dilute} binary granular mixtures.\cite{GBS21} As an application of this result, the stability analysis of the so-called homogeneous steady state (HSS) and the segregation problem driven by both a thermal gradient and gravity have been also studied in a subsequent paper. \cite{GBS24,GBS24a} For now, the study of binary mixtures has only considered the Boltzmann kinetic equation as the starting point and so, they are restricted to the low-density regime. Thus, it is quite natural to extend the analysis performed in Ref.\ \onlinecite{GBS21} to the (revised) Enskog kinetic
theory for a description of hydrodynamics and transport at higher densities.

Needless to say, the evaluation of the Navier--Stokes transport coefficients for a \emph{dense} granular mixture is quite an intricate problem, due mainly to the coupling between the different integral equations obeying the complete set of transport coefficients. Thus, to gain some insight into the general problem, we will make here a first approach to the description of a general mixture by considering a more simple situation: we consider a granular binary mixture where the concentration of one of the species (of mass $m_0$ and diameter $\sigma_0$) is very small (impurity or tracer limit). In this situation, one can assume that (i) the state of the dense granular gas (excess species of mass $m$ and diameter $\sigma$) is not perturbed by the presence of tracer particles, and (ii) one can also neglect collisions among tracer particles themselves in their corresponding kinetic equation. Under these conditions, the velocity distribution function $f$ of the granular gas obeys a (closed) nonlinear Enskog equation while the velocity distribution function $f_0$ of the tracer particles verifies a linear Enskog--Lorentz equation. At a kinetic level, the tracer limit greatly simplifies the application of the Chapman--Enskog method \cite{CC70} to a multicomponent granular mixture.

Since in the tracer limit the pressure tensor and the heat flux of the mixture (impurities or tracer particles plus granular gas) are the same as that for the excess species,\cite{GBS18} the mass transport of impurities $\mathbf{j}_0$ is the relevant flux of the problem. \vicente{In accordance with previous works on tracer diffusion in granular gases, \cite{GDH07,G08a,GV09}} 
the Navier–Stokes constitutive equation for the mass flux (that is, linear in the spatial gradients) can be written as
\beq
\label{0.1}
\mathbf{j}_0^{(1)}=-\frac{m_0^2}{\rho} D_0 \nabla n_0-\frac{m m_0}{\rho} D \nabla n-\frac{\rho}{T}D_T \nabla T,
\eeq
where $\rho=mn$ is the total mass density of the granular gas, $n_0$ is the number density of the tracer particles, $n$ is the number density of the gas particles, and $T$ is the granular temperature. In addition, $D_0$ is the tracer diffusion coefficient, $D$ is the mutual diffusion coefficient, and $D_T$ is the thermal diffusion coefficient. The determination of the diffusion transport coefficients $D_0$, $D$, and $D_T$ is the main objective of the present work. As for elastic collisions, \cite{CC70} these transport coefficients are given in terms of the solutions of a set of coupled linear integral equations. These integral equations are approximately solved by considering the leading terms in a Sonine polynomial expansion. However, as in the case of dilute granular mixtures, \cite{GBS21} evaluating the diffusion coefficients in the time-dependent problem requires numerically solving a set of nonlinear differential equations. To simplify the analysis, we focus here on the relevant state of confined systems with \emph{steady} granular temperature. This steady state allows for a more tractable approach and enables the derivation of analytical expressions for $D_0$, $D$, and $D_T$ in terms of the parameter space of the system.

To asses the accuracy of the (approximate) theoretical results, kinetic theory predictions for the tracer diffusion coefficient $D_0$ obtained in the first Sonine approximation are confronted against computer simulations carried out independently by both molecular dynamics (MD)\cite{TS05,AT17,FS23} and the direct simulation Monte Carlo (DSMC) method.\cite{B94} As in previous works, \cite{GM04,GV09} the diffusion coefficient is computed in the simulations from the mean-square displacement (MSD) of impurities immersed in a confined dense granular gas undergoing the HSS. 

Finally, since the explicit forms of the diffusion transport coefficients are at hand, a segregation criterion based on the thermal diffusion factor is derived. Segregation, induced by both gravity and a thermal gradient, shows the transition between the so-called Brazil-nut effect (BNE) and the reverse Brazil-nut effect (RBNE) by varying the parameters of the system. As expected, the corresponding phase diagrams characterizing the transition BNE/RBNE are different from those previously obtained  for inelastic hard spheres or disks (IHS).\cite{G08a,GV09}

The plan of the paper is as follows. The $\Delta$-model and the Enskog equation are introduced in Sec.\ \ref{sec2} for the granular gas while Sec.\ \ref{sec3} deals with the Enskog--Lorentz kinetic equation for the tracer particles. The application of the Chapman--Enskog method to solve the the Enskog--Lorentz equation to first order in spatial gradients is described in Sec.\ \ref{sec4}. The corresponding integral equations obeying the diffusion transport coefficients are also derived. Then, these integral equations are approximately solved up to the first Sonine approximation in Sec.\ \ref{sec5}. Some technical details of the calculations are provided in the Appendices \ref{appA} and \ref{appB}. The theoretical results for the tracer diffusion coefficient $D_0$ are compared with both DSMC results and MD simulations for several configurations in Sec.\ \ref{sec6}. Thermal diffusion segregation is analyzed in Sec.\ \ref{sec7}, while the paper is closed in Sec.\ \ref{sec8} with a brief discussion of the reported results.

\section{Enskog kinetic equation for a model of a confined quasi-two dimensional granular gas at moderate densities}
\label{sec2}

\subsection{Collision rules in the $\Delta$-model}

We consider a set of solid particles or grains of mass $m$ and diameter $\sigma$. Under rapid flow conditions, the set can be modeled as a gas of hard spheres with inelastic collisions. In the simplest case, the spheres are assumed to be completely smooth and hence, the inelasticity in collisions is characterized by the (positive) constant coefficient of normal restitution $\al$. As said in the Introduction, here we start from a collisional model (the $\Delta$-model) for a quasi-two dimensional geometry where grains are confined in the vertical direction. In this model, an extra velocity $\Delta$ (which is assumed to be constant) is added to the relative motion of the colliding spheres in such a way the magnitude of the normal component of the relative velocity is increased by a given factor in each binary collision. Thus, the relationship between the pre-collisional velocities $(\mathbf{v}_1, \mathbf{v}_2)$ and the post-collisional velocities $(\mathbf{v}_1',\mathbf{v}_2')$ in the $\Delta$-model is\cite{BRS13}
\beq
\label{1.1}
\mathbf{v}_1'=\mathbf{v}_1-\frac{1}{2}\left(1+\alpha\right)(\widehat{{\boldsymbol {\sigma }}}\cdot \mathbf{g}_{12})\widehat{{\boldsymbol {\sigma }}}-\Delta \widehat{{\boldsymbol {\sigma }}},
\eeq
\beq
\label{1.2}
{\bf v}_{2}'=\mathbf{v}_{2}+\frac{1}{2}\left(1+\alpha\right)
(\widehat{{\boldsymbol {\sigma}}}\cdot \mathbf{g}_{12})
\widehat{\boldsymbol {\sigma}}+\Delta \widehat{{\boldsymbol {\sigma }}},
\eeq
where $\mathbf{g}_{12}=\mathbf{v}_1-\mathbf{v}_2$ is the relative velocity and $\widehat{{\boldsymbol {\sigma}}}$ is the unit collision vector joining
the centers of the two colliding spheres and pointing from particle 1 to particle 2. Particles are approaching if $\widehat{{\boldsymbol {\sigma}}}\cdot \mathbf{g}_{12}>0$. In Eqs.\ \eqref{1.1} and \eqref{1.2}, $\al$ is the (constant) coefficient of normal restitution defined in the interval $0<\alpha\leq 1$, and $\Delta$ (which has dimensions of velocity) is positive. This extra velocity is compatible with angular momentum conservation \cite{L04bis} and  points outward in the normal direction $\widehat{\boldsymbol {\sigma}}$. From Eqs.\ \eqref{1.1} and  \eqref{1.2} is quite simple to get the relative velocity after collision $\mathbf{g}_{12}'$ as
\beq
\label{1.3}
\mathbf{g}_{12}'=\mathbf{v}_1'-\mathbf{v}_2'=\mathbf{g}_{12}-(1+\al)(\widehat{{\boldsymbol {\sigma}}}\cdot \mathbf{g}_{12})
\widehat{\boldsymbol {\sigma}}-2\Delta \widehat{{\boldsymbol {\sigma }}},
\eeq
so that
\beq
\label{1.4}
(\widehat{{\boldsymbol {\sigma}}}\cdot \mathbf{g}_{12}')=-\al (\widehat{{\boldsymbol {\sigma}}}\cdot \mathbf{g}_{12})-2\Delta.
\eeq

The collision rules \eqref{1.1} and \eqref{1.2} conserve momentum but energy is not conserved. The change in kinetic energy upon collision is
\beqa
\label{1.5}
\Delta E&\equiv& \frac{m}{2}\left(v_1^{'2}+v_2^{'2}-v_1^2-v_2^2\right)\nonumber\\
&=&m\left[\Delta^2+\al \Delta (\widehat{{\boldsymbol {\sigma}}}\cdot \mathbf{g}_{12})-\frac{1-\al^2}{4}(\widehat{{\boldsymbol {\sigma}}}\cdot \vicente{\mathbf{g}_{12})^2}\right].\nonumber\\
\eeqa
The right-hand side of Eq.\ \eqref{1.5} vanishes for elastic collisions ($\al=1$) and $\Delta=0$. According to Eq.\ \eqref{1.5},  $\Delta E>0$ (energy is gained in a binary collision) if $\widehat{{\boldsymbol {\sigma}}}\cdot \mathbf{g}_{12}<2\Delta /(1-\al)$ while $\Delta E<0$ (energy is lost in a binary collision) if $\widehat{{\boldsymbol {\sigma}}}\cdot \mathbf{g}_{12}>2\Delta /(1-\al)$.

For practical purposes, it is also convenient to consider the restituting collision $\left(\mathbf{v}_1'',\mathbf{v}_2''\right)\to \left(\mathbf{v}_1,\mathbf{v}_2\right)$ with the same collision vector $\widehat{{\boldsymbol {\sigma }}}$:
\beq
\label{1.6}
\mathbf{v}_1''=\mathbf{v}_1-\frac{1}{2}\left(1+\alpha^{-1}\right)(\widehat{{\boldsymbol {\sigma }}}\cdot \mathbf{g}_{12})\widehat{{\boldsymbol {\sigma }}}-\alpha^{-1}\Delta \widehat{{\boldsymbol {\sigma }}},
\eeq
\beq
\label{1.7}
\mathbf{v}_2''=\mathbf{v}_2+\frac{1}{2}\left(1+\alpha^{-1}\right)(\widehat{{\boldsymbol {\sigma }}}\cdot \mathbf{g}_{12})\widehat{{\boldsymbol {\sigma }}}+\alpha^{-1}\Delta \widehat{{\boldsymbol {\sigma }}}.
\eeq
\vicente{From Eqs.\ \eqref{1.6} and \eqref{1.7}, one gets the relationship
\beq
\label{1.7.1}
\mathbf{g}_{12}''=\mathbf{v}_1''-\mathbf{v}_2''=\mathbf{g}_{12}-(1+\al^{-1})(\widehat{{\boldsymbol {\sigma}}}\cdot \mathbf{g}_{12})
\widehat{\boldsymbol {\sigma}}-2\al^{-1}\Delta \widehat{{\boldsymbol {\sigma}}},
\eeq
so that
\beq
\label{1.7.2}
(\widehat{{\boldsymbol {\sigma}}}\cdot \mathbf{g}_{12}'')=-\al^{-1} (\widehat{{\boldsymbol {\sigma}}}\cdot \mathbf{g}_{12})-2\al^{-1}\Delta.
\eeq
}
In addition, the volume transformation in velocity space  in a direct collision is $d \mathbf{v}_1'd \mathbf{v}_2'=\al d\mathbf{v}_1 d\mathbf{v}_2$, and so
$d \mathbf{v}_1'' d \mathbf{v}_2''=\al^{-1} d \mathbf{v}_1 d \mathbf{v}_2$.

\subsection{Enskog kinetic equation}

We assume that the granular gas is in the presence of a gravitational field $\mathbf{g}$.
At a kinetic level, all the relevant information on the state of the system is provided by the knowledge of the one-particle velocity distribution function $f(\mathbf{r}, \mathbf{v}; t)$.
For moderate densities, this distribution verifies the Enskog kinetic equation \cite{BGMB13,BRS13}
\beq
\label{1.8}
\frac{\partial f}{\partial t}+\mathbf{v}\cdot \nabla f+\mathbf{g}\cdot \frac{\partial f}{\partial \mathbf{v}}=J[\mathbf{r},\mathbf{v}|f,f],
\eeq
where the Enskog collision operator $J$ of the $\Delta$-model reads
\begin{widetext}
\beqa
\label{1.9}
J[\mathbf{r},\mathbf{v}_1|f,f]&\equiv& \sigma^{d-1}\int d{\bf v}_{2}\int d\widehat{\boldsymbol{\sigma}}
\Theta (-\widehat{{\boldsymbol {\sigma }}}\cdot {\bf g}_{12}-2\Delta)
(-\widehat{\boldsymbol {\sigma }}\cdot {\bf g}_{12}-2\Delta)
\al^{-2}\chi(\mathbf{r},\mathbf{r}+\boldsymbol{\sigma}) f(\mathbf{r},\mathbf{v}_1'';t)\nonumber\\
& & \times
f(\mathbf{r}+\boldsymbol{\sigma},\mathbf{v}_2'';t)
-\sigma^{d-1}\int\ d{\bf v}_{2}\int d\widehat{\boldsymbol{\sigma}}
\Theta (\widehat{{\boldsymbol {\sigma }}}\cdot {\bf g}_{12})
(\widehat{\boldsymbol {\sigma }}\cdot {\bf g}_{12})
\chi(\mathbf{r},\mathbf{r}+\boldsymbol{\sigma}) f(\mathbf{r},\mathbf{v}_1;t)
f(\mathbf{r}+\boldsymbol{\sigma},\mathbf{v}_2;t).
\eeqa
\end{widetext}
\vicente{As in the case of the conventional IHS model, \cite{BP04,G19} the quantity $(-\widehat{\boldsymbol {\sigma }}\cdot {\bf g}_{12}-2\Delta)
\al^{-2}$ appearing in the gain term of the Enskog collision operator \eqref{1.9} arises from the length of the collision cylinder and the Jacobian of the transformation $(\mathbf{v}_1'',\mathbf{v}_2'')\to (\mathbf{v}_1,\mathbf{v}_2)$
} Like the Boltzmann equation, the Enskog equation neglects velocity correlations among particles that are about to collide, but it accounts for excluded volume effects through the pair correlation function $\chi(\mathbf{r},\mathbf{r}\pm\boldsymbol{\sigma})$. In Eq.\ \eqref{1.9},  $\Theta(x)$ is the Heaviside step function and  $d$ is the dimensionality of the system ($d=2$ for disks and $d=3$ for spheres). Note that although the system considered is two-dimensional, the calculations worked out here will be performed by an arbitrary number of dimensions $d$.

An important property of the Enskog collision operator is \cite{BGMB13,SRB14}
\begin{widetext}
\beqa
\label{1.10}
I_\psi&\equiv& \int\; d\mathbf{v}_1\; \psi(\mathbf{v}_1) J[\mathbf{r},\mathbf{v}_1|f,f]\nonumber\\
&=&\sigma^{d-1}\int \;d\mathbf{v}_1\int\ d{\bf v}_{2}\int d\widehat{\boldsymbol{\sigma}}\,
\Theta (\widehat{{\boldsymbol {\sigma }}}\cdot {\bf g}_{12})(\widehat{\boldsymbol {\sigma }}\cdot {\bf g}_{12})
\chi(\mathbf{r},\mathbf{r}+\boldsymbol{\sigma}) f(\mathbf{r},\mathbf{v}_1;t)
f(\mathbf{r}+\boldsymbol{\sigma},\mathbf{v}_2;t)
\left[\psi(\mathbf{v}_1')-\psi(\mathbf{v}_1)\right],
\nonumber\\
\eeqa
\end{widetext}
where $\mathbf{v}_1'$ is defined by Eq.\ \eqref{1.1}. The property \eqref{1.10} is the same as for the conventional model of IHS. A consequence of the relation \eqref{1.10} is that the balance equations of the densities of mass, momentum, and energy can be derived by following similar mathematical steps as those made for IHS. They are given by
\cite{GBS18}
\begin{equation}
\label{1.11}
D_t n+n\nabla \cdot \mathbf{U}=0,
\end{equation}
\begin{equation}
\label{1.12}
\rho D_t U_i+\partial_j P_{ij}=\rho \mathbf{g},
\end{equation}
\begin{equation}
\label{1.13}
D_t T+\frac{2}{dn}\left(\partial_i q_i+ P_{ij}\partial_j U_i \right)=-\zeta T,
\end{equation}
where $D_t\equiv \partial_t+\mathbf{U}\cdot \nabla$ is the material derivative, $\partial_i\equiv \partial/\partial r_i$, and $\rho=mn$ is the mass density. In terms of the distribution $f$, the hydrodynamic fields $n$, $\mathbf{U}$, and $T$ are defined as
\beq
\label{1.13.0}
\left\{n, n\mathbf{U}, d n T\right\}=\int d\mathbf{v}\; \left\{1, \mathbf{v}, m V^2\right\}f(\mathbf{v}),
\eeq
\vicente{where 
\beq
\label{1.13.1a}
\mathbf{V}=\mathbf{v}-\mathbf{U}
\eeq
}
is the peculiar velocity. The forms of the pressure tensor $P_{ij}$, the heat flux $\mathbf{q}$, and the cooling rate $\zeta$ in terms of the distribution $f$ can be found in Ref.\ \onlinecite{GBS18}. In addition, the Enskog equation \eqref{1.8} has been also solved by means of the Chapman--Enskog method \cite{CC70} and explicit expressions for the Navier--Stokes transport coefficients and the cooling rate have been derived by assuming steady state conditions. \cite{GBS18,GBS20}

\subsection{Homogeneous steady state}

We consider the HSS in the absence of the gravitational field ($\mathbf{g}=\mathbf{0}$). In this simple situation, the heat flux vanishes ($\mathbf{q}=\mathbf{0}$) and the pressure tensor $P_{ij}= p \delta_{ij}$, where $p$ is the hydrostatic pressure. \vicente{It is given by $p=nT p^*$, where \cite{GBS18}
\beq
\label{1.13.1b}
p^*=1+2^{d-2}\chi \phi (1+\al)+\frac{2^d}{\sqrt{2\pi}}\chi \phi \Delta^*.
\eeq
}
In Eq.\ \eqref{1.13.1b}, the solid volume fraction $\phi$ is defined as
\beq
\label{1.14}
\phi=\frac{\pi^{d/2}}{2^{d-1}d\Gamma\left(\frac{d}{2}\right)}n \sigma^d,
\eeq
while $\Delta^*=\Delta/v_0$, with $v_0=\sqrt{2T/m}$ being the thermal velocity. Note that here $T$ means the steady state value of the granular temperature.
\RB{The quantity $\Delta^*$ is a dimensionless parameter
as the stationary temperature $T$ is proportional to $\Delta^2$ (see Ref.~\onlinecite{SRB14}). }

\vicente{
In order to get the relationship between $\Delta^*$ and $\al$, one has to give the form of the pair correlation function $\chi$. In the case of a two-dimensional system ($d=2$), a good approximation for $\chi$ is \cite{JM87}
\begin{equation}
\label{4.11}
\chi=\frac{1-\frac{7}{16}\phi}{(1-\phi)^2}.
\end{equation}
}

\vicente{Moreover, the velocity distribution function $f(\mathbf{v})$ is needed to determine the cooling rate $\zeta$. On the other hand, the exact form of $f(\mathbf{v})$ in the HSS is not known to date. An indirect information on this distribution is given through its fourth-degree moment or \textit{kurtosis} $a_2$; this quantity measures the departure of the true distribution from its Maxwellian form. Previous results of the $\Delta$-model \cite{BMGB14} have clearly shown that the magnitude of $a_2$ is in general very small in the HSS (see Fig.\ 1 of Ref.\ \onlinecite{GBS18}). Thus, a good estimate of the cooling rate $\zeta$ can be achieved by replacing $f$ by the Maxwellian distribution   } 
\beq
\label{1.15}
f(\mathbf{v})\to f_{\text{M}}(\mathbf{v})=n \left(\frac{m}{2\pi T}\right)^{d/2} \exp \left(-\frac{m v^2}{2T}\right).
\eeq
Using the approximation \eqref{1.15}, $\zeta$ can be written as
\beq
\label{1.16}
\zeta=\frac{\sqrt{2}\pi^{\frac{d-1}{2}}}{d\Gamma\left(\frac{d}{2}\right)}n\sigma^{d-1} v_0 \chi(\phi) \left(1-\al^2-2\Delta^{*2}-\sqrt{2\pi}\al \Delta^*\right).
\eeq
In the HSS, $\partial_t T=0$ and so Eq.\ \eqref{1.13} implies that the cooling rate $\zeta=0$. The condition $\zeta=0$ yields a quadratic equation in $\Delta^*$, whose physical solution is
\beq
\label{1.17}
\Delta^*(\al)=\frac{1}{2}\sqrt{\frac{\pi}{2}}\al\left[\sqrt{1+
\frac{4(1-\al^2)}{\pi \al^2}}-1\right].
\eeq
\RS{Recalling that $\Delta^*=\Delta/\sqrt{2T/m}$, Eq.~\eqref{1.17} gives the value of the stationary temperature in terms of the imposed values of $\alpha$ and $\Delta$. Equivalently, in this dimensionless form, this result indicates that the temperature evolves such that at the HSS $\Delta^*$ is given by Eq.~\eqref{1.17}.}
As expected, one concludes that the steady state for elastic particles is only achieved for a vanishing extra velocity, that is, $\Delta^*(1)=0$.
\RS{Using MD simulations, it has been shown that the relation \eqref{1.17} remains accurate, with deviations smaller than 2\%, except for high densities and inelasticities.~\cite{BRS13}}

\RS{Equation~\eqref{1.17} predicts that in the HSS, the stationary temperature diverges when $\alpha\to 1$ if $\Delta$ is set fixed. This result has been verified in MD simulations of the $\Delta$-model (see Fig.~2 of Ref.~\onlinecite{BRS13}). The same divergence has been observed in MD of a three-dimensional system with vibrating walls, with the stationary temperature scaling as the wall velocity squared with a prefactor that depends on the box height (see Fig.~4 of Ref.~\onlinecite{MGB19}). 
\RB{Moreover, there is a qualitative agreement between the stationary temperature $T$ obtained in MD simulations\cite{MGB19} and its theoretical prediction provided by the $\Delta$-model. \cite{SRB14}} Such agreement validates the assumption of the $\Delta$-model of considering a fixed value of $\Delta$ and suggests that its value should depend on the vertical confinement and scales with the wall velocity.}

\section{Impurities immersed in a confined granular gas}
\label{sec3}

We suppose now that a few tracer or impurity particles of mass $m_0$ and diameter $\sigma_0$ are added to the granular gas. Since the concentration of the tracer species is negligibly small, their presence does not perturb the state of the granular gas. This means that the velocity distribution function $f(\mathbf{r}, \mathbf{v}; t)$ still verifies the Enskog equation \eqref{1.8} and so, the balance equations for the number density $n$, the mean flow velocity $\mathbf{U}$, and the temperature $T$ for the granular gas are given by Eqs.\ \eqref{1.11}, \eqref{1.12}, and \eqref{1.13}, respectively. In addition, as mentioned in Sec.\ \ref{sec1}, in this paper we are mainly interested
in the evaluation of the transport coefficients defining the mass flux of the impurities since in the tracer limit the pressure tensor and the heat flux are the same as those of the granular gas (excess component).

Under the above conditions, the velocity distribution function $f_0(\mathbf{r}, \mathbf{v}; t)$ of the tracer particles satisfies the Enskog--Lorentz kinetic equation
\beq
\label{2.1}
\frac{\partial f_0}{\partial t}+\mathbf{v}\cdot \nabla f_0+\mathbf{g}\cdot \frac{\partial f_0}{\partial \mathbf{v}}=J_0[\mathbf{r},\mathbf{v}|f,f_0],
\eeq
where the Enskog--Lorentz  collision operator $J_0$ of the $\Delta$-model reads \cite{GBS21}
\begin{widetext}
\beqa
\label{2.2}
J_0[\mathbf{r},\mathbf{v}_1|f,f]&\equiv& \bar{\sigma}^{d-1}\int d{\bf v}_{2}\int d\widehat{\boldsymbol{\sigma}}
\Theta (-\widehat{{\boldsymbol {\sigma }}}\cdot {\bf g}_{12}-2\Delta_0)
(-\widehat{\boldsymbol {\sigma }}\cdot {\bf g}_{12}-2\Delta_0)
\al_0^{-2}\chi_0(\mathbf{r},\mathbf{r}+\bar{\boldsymbol{\sigma}}) f_0(\mathbf{r},\mathbf{v}_1'';t)\nonumber\\
& & \times
f(\mathbf{r}+\bar{\boldsymbol{\sigma}},\mathbf{v}_2'';t)
-\bar{\sigma}^{d-1}\int\ d{\bf v}_{2}\int d\widehat{\boldsymbol{\sigma}}
\Theta (\widehat{{\boldsymbol {\sigma }}}\cdot {\bf g}_{12})
(\widehat{\boldsymbol {\sigma }}\cdot {\bf g}_{12})
\chi_0(\mathbf{r},\mathbf{r}+\bar{\boldsymbol{\sigma}}) f_0(\mathbf{r},\mathbf{v}_1;t)
f(\mathbf{r}+\boldsymbol{\sigma},\mathbf{v}_2;t).
\nonumber\\
\eeqa
\end{widetext}
Here, $\bar{\boldsymbol{\sigma}}=\bar{\sigma}\widehat{\boldsymbol{\sigma}}$, $\bar{\sigma}=(\sigma_0+\sigma)/2$, $\al_0\leq 1$ is the (positive) coefficient of restitution for impurity-gas collisions, and $\chi_0$ is the pair correlation function for impurity-gas pairs at contact. Note also that in Eq.\ \eqref{2.1} the mutual interactions among the impurity particles have been neglected as compared with their interactions with the particles of the granular gas. In Eq.\ \eqref{2.2}, the relationship between $(\mathbf{v}_1'',\mathbf{v}_2'')$ and $(\mathbf{v}_1,\mathbf{v}_2)$ is
\beq
\label{2.2.1}
\mathbf{v}_1''=\mathbf{v}_1-M\left(1+\alpha_{0}^{-1}\right)(\widehat{{\boldsymbol {\sigma }}}\cdot \mathbf{g}_{12})\widehat{{\boldsymbol {\sigma }}}-2M\Delta_{0}\al_{0}^{-1} \widehat{{\boldsymbol {\sigma }}},
\eeq
\beq
\label{2.2.2}
\mathbf{v}_2''=\mathbf{v}_2+M_0\left(1+\alpha_{0}^{-1}\right)(\widehat{{\boldsymbol {\sigma }}}\cdot \mathbf{g}_{12})\widehat{{\boldsymbol {\sigma}}}+2M_0\Delta_{0}\al_{0}^{-1} \widehat{{\boldsymbol {\sigma }}},
\eeq
where
\beq
\label{2.2.3}
M=\frac{m}{m+m_0}, \quad M_0=\frac{m_0}{m+m_0}.
\eeq
Equations \eqref{2.2.1}--\eqref{2.2.2} yield the relationship
\beq
\label{2.2.4}
(\widehat{{\boldsymbol {\sigma}}}\cdot \mathbf{g}_{12}'')=-\al_{0}^{-1}(\widehat{{\boldsymbol {\sigma}}}\cdot \mathbf{g}_{12})-2\Delta_{0}\al_{0}^{-1},
\eeq
where $\mathbf{g}_{12}''=\mathbf{v}_1''-\mathbf{v}_2''$.

Similarly, the collision rules for the direct collisions $(\mathbf{v}_1,\mathbf{v}_2)\to (\mathbf{v}_1',\mathbf{v}_2')$ with the same collision vector $\widehat{{\boldsymbol {\sigma}}}$ are defined as 
\beq
\label{2.2.1bis}
\mathbf{v}_1'=\mathbf{v}_1-M\left(1+\alpha_{0}\right)(\widehat{{\boldsymbol {\sigma }}}\cdot \mathbf{g}_{12})\widehat{{\boldsymbol {\sigma }}}-2M\Delta_{0}\widehat{{\boldsymbol {\sigma }}},
\eeq
\beq
\label{2.2.2bis}
\mathbf{v}_2'=\mathbf{v}_2+M_0\left(1+\alpha_{0}\right)(\widehat{{\boldsymbol {\sigma }}}\cdot \mathbf{g}_{12})\widehat{{\boldsymbol {\sigma}}}+2M_0\Delta_{0} \widehat{{\boldsymbol {\sigma }}}.
\eeq
Equations \eqref{2.2.1bis}--\eqref{2.2.2bis} lead to the relationship
\beq
\label{2.2.4bis}
(\widehat{{\boldsymbol {\sigma}}}\cdot \mathbf{g}_{12}')=-\al_{0}(\widehat{{\boldsymbol {\sigma}}}\cdot \mathbf{g}_{12})-2\Delta_{0},
\eeq
where $\mathbf{g}_{12}'=\mathbf{v}_1'-\mathbf{v}_2'$.

The number density for the impurities is defined as
\begin{equation}
\label{2.3}
n_0({\bf r},t)=\int\; d{\bf v}f_0({\bf r},{\bf v},t).
\end{equation}
The impurities may freely loose or gain momentum and energy in its interactions with the particles of the granular gas. Consequently, the momentum and energy are not
invariants of the collision operator $J_{0}[{\bf v}|f_0,f]$. Only the number density $n_0$ is conserved; its
continuity equation can be directly obtained from Eq.\ (\ref{2.1}) as
\begin{equation}
\label{2.4}
D_{t}n_0+n_0\nabla \cdot {\bf U}+\frac{\nabla \cdot {\bf j}_0}{m_0}=0\;,
\end{equation}
where
\begin{equation}
{\bf j}_{0}=m_{0}\int d{\bf v}\,{\bf V}\,f_0({\bf r},{\bf v},t)
\label{2.5}
\end{equation}
is the mass flux for the impurities, relative to the local flow ${\bf U}$. Although the granular temperature $T$ is the relevant one at a hydrodynamic level, an interesting quantity at a kinetic level is the local temperature of the \vicente{impurities (or tracer particles)} $T_0$. This quantity measures the mean kinetic energy of the impurities. It is defined as
\begin{equation}
\label{2.6}
T_0({\bf r}, t)=\frac{m_0}{d n_0({\bf r}, t)}\int \; d{\bf v}\, V^2 f_0({\bf r},{\bf v},t).
\end{equation}
As confirmed by kinetic theory calculations and computer simulations, \cite{BSG20} the global temperature $T$ and the temperature of impurities $T_0$ are in general different. This means that the granular energy per particle is not equally distributed between both species of the mixture.

\subsection{Homogeneous steady state for the impurities}

Before considering inhomogeneous states for the impurities, it is convenient to characterize the HSS. This state has been widely analyzed in Ref.\ \onlinecite{BSG20} by theoretical approaches and computer simulations. Since in the steady state $\partial_t T_0=0$, then [analogously to Eq. \ \eqref{1.13}] the   temperature ratio $T_0/T$ is determined from the condition $\zeta_0=0$. Here, $\zeta_0$ is the cooling rate for the impurities in the HSS. \vicente{As for the granular gas, the velocity distribution function $f_0(\mathbf{v})$ of the impurities is not exactly known. However, as occurs in the calculation of the global cooling rate $\zeta$, a good estimate of $\zeta_0$ can be achieved by replacing $f$ by $f_\text{M}$ and $f_0$ by the Maxwellian distribution}
\beq
\label{2.6.1}
f_0(\mathbf{v})\to f_{0,\text{M}}(\mathbf{v})=n_0 \left(\frac{m_0}{2\pi T_0}\right)^{d/2} \exp \left(-\frac{m_0 v^2}{2T_0}\right).
\eeq
In this approximation, $\zeta_0=\nu \zeta_0^*$ where $\nu=n\sigma^{d-1} v_0$ and $\zeta_0^*$ is \cite{BSG20}
\begin{widetext}
\beqa
\label{2.6.2}
\zeta_0^*&=&\frac{4\pi^{(d-1)/2}}{d\Gamma\left(\frac{d}{2}\right)}
\left(\frac{\bar{\sigma}}{\sigma}\right)^{d-1}
M \chi_0 (1+\al_{0})\theta^{-1/2}\left(1+\theta\right)^{1/2}
\left[1-\frac{1}{2}M(1+\alpha_{0})(1+\theta) \right]\nonumber\\
& &-\frac{4\pi^{d/2}}{d\Gamma\left(\frac{d}{2}\right)}\left(\frac{\bar{\sigma}}{\sigma}\right)^{d-1}M \chi_0\Delta_{0}^*\left[
\frac{2M\Delta_{0}^*}{\sqrt{\pi}}\theta^{1/2}\left(1+\theta\right)^{1/2}
-1+M(1+\al_{0})\left(1+\theta\right)\right].
\eeqa
\end{widetext}
Here, $\Delta_0^*=\Delta_0/v_0$ and $\theta=m_0 T/m T_0$ is the ratio between the mean square velocities of the impurity and particles of the granular gas. \vicente{The accuracy of the estimate \eqref{2.6.2} is justified by the good agreement found between the theoretical predictions of the temperature ratio $T_0/T$ (which are based on Eq.\ \eqref{2.6.2}) and computer simulations. \cite{BSG20} 
}

\vicente{To obtain the dependence of the temperature ratio on the parameter space of the system, one needs to give the (approximate) form of $\chi_0$. As in the case of $\chi$, a good approximation for $\chi_0$ for hard disks ($d=2$) is \cite{JM87} 
\begin{equation}
\label{4.12}
\chi_0=\frac{1}{1-\phi}+\frac{9}{8}\frac{\omega}{1+\omega}\frac{\phi}{(1-\phi)^2}.
\end{equation}
}

\section{Diffusion transport coefficients}
\label{sec4}

\vicente{
The Chapman--Enskog method \cite{CC70} is applied in this section to solve the Enskog--Lorentz equation \eqref{2.1} up to first order in spatial gradients. As widely discussed in many textbooks, \cite{CC70,FK72} there are two different stages in the relaxation of a molecular gas toward equilibrium. For times of the order of the mean free time, a kinetic stage is first identified where the effect of collisions is to relax the gas toward the so-called \textit{local} equilibrium state. Then, a slow stage referred to as the \textit{hydrodynamic} regime is reached where the system has forgotten its initial conditions. The main feature of this regime is that the microscopic state of the granular gas is governed by the hydrodynamic fields (in the case of a binary mixture by $n_0$, $n$, $\mathbf{U}$, and $T$). The above two stages are also expected in the case of granular gases except that in the kinetic stage the distribution function will generally relax toward a time-dependent nonequilibrium distribution (the homogeneous cooling state in the conventional IHS model) instead of the local equilibrium distribution. A crucial point is that although the granular temperature $T$ is not a conserved field (due to the inelastic character of the collisions), it is assumed that $T$ can still be considered as a slow field. This assumption has been clearly supported by the good agreement found between granular hydrodynamics and computer simulations in several non-equilibrium situations. \cite{DHGD02,G19}   More details on the application of the Chapman--Enskog method to granular mixtures can be found for example in Ref.\ \onlinecite{G19}.

Based on the above arguments}, in the hydrodynamic regime, the Enskog--Lorentz equation \eqref{2.1} admits a \emph{normal} (or hydrodynamic) solution where all the space and time dependence of $f_0$ only occurs through a functional dependence on the hydrodynamic fields. As usual, \cite{CC70} this functional dependence can be made explicit by assuming small spatial gradients. In this case, $f_0$ can be written as a series expansion in powers of the spatial gradients of the hydrodynamic fields:
\beq
\label{3.1}
f_0=f_0^{(0)}+f_0^{(1)}+\cdots,
\eeq
where the approximation $f_0^{(k)}$ is of order $k$ in the spatial gradients. The implementation of the Chapman--Enskog method to first order in the spatial gradients follows similar steps as those made in the conventional IHS model. \cite{GD02,GDH07,GHD07,GV09} Here, only the final results for the integral equations verifying the diffusion transport coefficients will be displayed.

The first-order distribution function $f_0^{(1)}(\mathbf{V})$ is given by
\beqa
\label{3.2}
f_0^{(1)}&=&\vicente{\boldsymbol{\mathcal{A}}_{0}\cdot \nabla T+\boldsymbol{\mathcal{B}}_{0}\cdot \nabla n_0+\boldsymbol{\mathcal{C}}_{0}\cdot \nabla n}+\mathcal{D}_{0,ij}\nabla_i U_j\nonumber\\
& &+\mathcal{E}_0 \nabla \cdot \mathbf{U},
\eeqa
where an implicit summation over repeated indices is used in Eq.\ \eqref{3.2}. \vicente{Note that the first-order distribution $f_0^{(1)}$ is written in terms of the spatial gradients of the hydrodynamic fields. 
\RG{According to the explanation given at the beginning of the section, the granular temperature $T$ is considered as a slow variable. In contrast, the partial temperature $T_0$ is a kinetic quantity that evolves more rapidly and its evolution is significantly influenced by the granular temperature $T$. Therefore, there is no term proportional to $\nabla T_0$ in the expression for $f_0^{(1)}$. } }

The quantities $\boldsymbol{\mathcal{A}}_{0}(\mathbf{V})$, $\boldsymbol{\mathcal{B}}_{0}(\mathbf{V})$, $\boldsymbol{\mathcal{C}}_{0}(\mathbf{V})$, $\mathcal{D}_{0,ij}(\mathbf{V})$, and $\mathcal{E}_{0}(\mathbf{V})$ obey certain integral equations. Since we are interested here in obtaining the diffusion transport coefficients, we will only pay attention to the unknowns $\boldsymbol{\mathcal{A}}_{0}(\mathbf{V})$, $\boldsymbol{\mathcal{B}}_{0}(\mathbf{V})$, and  $\boldsymbol{\mathcal{C}}_{0}(\mathbf{V})$ (this is equivalent to assume a nonequilibrium state with vanishing flow velocity).  

The first-order contribution $\mathbf{j}_0^{(1)}$ to the mass flux is given by Eq.\ \eqref{0.1}. In terms of $\boldsymbol{\mathcal{A}}_{0}$, $\boldsymbol{\mathcal{B}}_{0}$, and  $\boldsymbol{\mathcal{C}}_{0}$, the diffusion transport coefficients $D_T$, $D_0$, and $D$ are defined, respectively, as
\begin{equation}
D_{T}=-\frac{m_0}{\rho d}\int d\mathbf{v}\; \mathbf{V}\cdot \boldsymbol{\mathcal{A}}_{0}\left(
\mathbf{V}\right). \label{3.3}
\end{equation}
\begin{equation}
D_{0}=-\frac{\rho}{m_{0}n_{0}d}\int d\mathbf{v}\; \mathbf{v}\cdot \boldsymbol{\mathcal{B}}_{0}\left(
\mathbf{V}\right) ,\label{3.4}
\end{equation}
\begin{equation}
D=-\frac{1}{d}\int d\mathbf{v}\; \mathbf{V}\cdot \boldsymbol{\mathcal{C}}_{0}\left( \mathbf{V}\right),
\label{3.5}
\end{equation}

The unknowns $\boldsymbol{\mathcal{A}}_{0}\left(\mathbf{V}\right)$, $\boldsymbol{\mathcal{B}}_{0}\left(\mathbf{V}\right)$, and $\boldsymbol{\mathcal{C}}_{0}\left(\mathbf{V}\right)$ are the solutions of the following set of coupled linear integral equations:
\beqa
\label{3.6}
& & -\zeta^{(0)} T\partial_T \boldsymbol{\mathcal{A}}_{0}-\frac{1}{2}\zeta^{(0)}\Big(1-\Delta^*\frac{\partial \ln \zeta^*}
{\partial \Delta^*}\Big)\boldsymbol{\mathcal{A}}_{0}-J_0^{(0)}[\boldsymbol{\mathcal{A}}_{0},f^{(0)}]\nonumber\\
& & =\mathbf{A}_0+
J_0^{(0)}[f_0^{(0)},\boldsymbol{\mathcal{A}}],
\eeqa
\beq
\label{3.7}
-\zeta^{(0)} T\partial_T \boldsymbol{\mathcal{B}}_{0}-J_0^{(0)}[\boldsymbol{\mathcal{B}}_{0},f^{(0)}]=\mathbf{B}_0,
\eeq
\beqa
\label{3.8}
& & -\zeta^{(0)} T\partial_T \boldsymbol{\mathcal{C}}_{0}-J_0^{(0)}[\boldsymbol{\mathcal{C}}_{0},f^{(0)}]=\mathbf{C}_0+\zeta^{(0)}
\Big(1+\phi \frac{\partial \ln \chi}{\partial \phi}\Big)\nonumber\\
& & \times \boldsymbol{\mathcal{A}}_{0}+
J_0^{(0)}[f_0^{(0)},\boldsymbol{\mathcal{C}}].
\eeqa
In Eqs.\ \eqref{3.6}--\eqref{3.8}, $\zeta^{(0)}$ is the zeroth-order contribution to $\zeta$ and $\zeta^*=\zeta^{(0)}/\nu$. An explicit (but approximate) form of the cooling rate $\zeta^{(0)}$ is given by Eq.\ \eqref{1.16}. In addition, the coefficients $\mathbf{A}_0$, $\mathbf{B}_0$, and $\mathbf{C}_0$ are given, respectively, by
\beqa
\label{3.9}
\mathbf{A}_0(\mathbf{V})&=&-\mathbf{V}T \frac{\partial f_0^{(0)}}{\partial T}-\frac{p}{\rho}\Big(1-\frac{1}{2}\Delta^*\frac{\partial \ln p^*}{\partial \Delta^*}\Big) \frac{\partial f_0^{(0)}}{\partial \mathbf{V}}\nonumber\\
& & -\boldsymbol{\mathcal{K}}_0\Big[T \frac{\partial f^{(0)}}{\partial T}\Big],
\eeqa
\beq
\label{3.10}
\mathbf{B}_0(\mathbf{V})=-\mathbf{V}n_0 \frac{\partial f_0^{(0)}}{\partial n_0},
\eeq
\beqa
\label{3.11}
\mathbf{C}_0(\mathbf{V})&=&-\mathbf{V}n \frac{\partial f_0^{(0)}}{\partial n}-m^{-1}\frac{\partial p}{\partial n} \frac{\partial f_0^{(0)}}{\partial \mathbf{V}}\nonumber\\
& & -\frac{(1+\omega)^{-d}}{\chi_0 T}\Big(\frac{\partial \mu_0}{\partial \phi}\Big)_{T,n_0}\boldsymbol{\mathcal{K}}_0\Big[f^{(0)}\Big].
\eeqa
Here, $\omega= \sigma_0/\sigma$ is the diameter ratio, $\mu_0$ is the chemical potential of the impurity, and the operator $\boldsymbol{\mathcal{K}}_0\Big[X\Big]$ is defined as
\begin{widetext}
\beqa
\label{2.21}
\boldsymbol{\mathcal{K}}_0\Big[X\Big]&=&-\bar{\sigma}^{d}\chi_0 \int d{\bf v}_{2}\int d\widehat{\boldsymbol{\sigma}}
\Theta (-\widehat{{\boldsymbol {\sigma }}}\cdot {\bf g}_{12}-2\Delta_0)(-\widehat{{\boldsymbol {\sigma }}}\cdot {\bf g}_{12}-2\Delta_0)
\widehat{\boldsymbol{\sigma}}\al_0^{-2}f_0^{(0)}(\mathbf{V}_1'')X(\mathbf{V}_2'')\nonumber\\
& &
+\bar{\sigma}^{d}\chi_0 \int d{\bf v}_{2}\int d\widehat{\boldsymbol{\sigma}}
\Theta (\widehat{{\boldsymbol {\sigma}}}\cdot {\bf g}_{12})(\widehat{{\boldsymbol {\sigma}}}\cdot {\bf g}_{12})
\widehat{\boldsymbol{\sigma}}f_0^{(0)}(\mathbf{V}_1) X(\mathbf{V}_2).
\eeqa
In addition,
\beqa
\label{2.21.1}
J_0^{(0)}[f_0^{(0)},X]&=&
\bar{\sigma}^{d-1}\chi_0 \int d{\bf v}_{2}\int d\widehat{\boldsymbol{\sigma}}
\Theta (-\widehat{{\boldsymbol {\sigma }}}\cdot {\bf g}_{12}-2\Delta_0)
(-\widehat{\boldsymbol {\sigma }}\cdot {\bf g}_{12}-2\Delta_0)
\al_0^{-2}f_0^{(0)}(\mathbf{V}_1'')X(\mathbf{V}_2'')\nonumber\\
& &
-\bar{\sigma}^{d-1}\chi_0\int\ d{\bf v}_{2}\int d\widehat{\boldsymbol{\sigma}}
\Theta (\widehat{{\boldsymbol {\sigma }}}\cdot {\bf g}_{12})
(\widehat{\boldsymbol {\sigma }}\cdot {\bf g}_{12})f_0^{(0)}(\mathbf{V}_1)
X(\mathbf{V}_2),
\eeqa
and we have accounted for that the first-order distribution $f^{(1)}(\mathbf{V})$ of the granular gas is 
\beq
\label{2.22}
f^{(1)}(\mathbf{V})=\boldsymbol{\mathcal{A}}(\mathbf{V})\cdot \nabla \ln T+\boldsymbol{\mathcal{C}}(\mathbf{V})\cdot \nabla \ln n+\mathcal{D}_{ij}(\mathbf{V})\nabla_i U_j+E(\mathbf{V}) \nabla \cdot \mathbf{U},
\eeq
where the quantities $\boldsymbol{\mathcal{A}}$, $\boldsymbol{\mathcal{C}}$, $\mathcal{D}_{ij}$ and $E$ obey a set of coupled linear integral equations.\footnote{The term $-(p/\rho)(1+T\partial_T\ln p^*)(\partial f^{(0)}/\partial \mathbf{V})$
was neglected in the expression (47) of the quantity $\mathbf{A}(\mathbf{V})$ derived in Ref.\ \onlinecite{GBS18}. This term could give rise to a nonzero contribution to the kinetic thermal conductivity $\kappa_\text{k}$ of the granular gas. However, when one takes the Maxwellian approximation to the zeroth-order distribution $f^{(0)}$, this contribution to $\kappa_\text{k}$ vanishes. Thus, within this approximation, the expressions of the Navier--Stokes transport coefficients obtained in Ref.\ \onlinecite{GBS18} remain unaltered.}
\end{widetext}

\vicente{As mentioned in several previous works, \cite{GDH07,G11} the explicit form of the diffusion coefficient $D$ requires the knowledge of the functional derivative of the (local) pair distribution function $\chi_0$ with respect to the density $n$ (see for instance Eq.\ (C.11) of Ref.\ \onlinecite{GDH07}).  However, in view of the mathematical difficulties involved in evaluating this functional derivative for granular gases, it is usually determined by requiring that the expressions of the diffusion transport coefficients for granular mixtures \cite{GHD07} reduce for elastic collisions to those obtained many years ago by L\'opez de Haro \textit{et al.} \cite{LCK83} in the \textit{revised} Enskog theory for molecular mixtures of hard spheres. This is the reason for which the chemical potential $\mu_0$ is present in our theory (see Eq.\ \eqref{3.11}). 

Since granular gases are systems which are inherently in nonequilibrium conditions, they lack a thermodynamic description, and thus the use of the chemical potential concept might be questionable. However, as mentioned before, the presence of the term $\partial \mu_0/\partial \phi$ in Eq.\ \eqref{3.11} results from the functional derivative of $\chi_0$ with respect to the gas density $n$. Here, as in previous works, \cite{G11} for practical purposes the expression used for $\mu_0$ is the same as that obtained for an ordinary gas mixture ($\al_0=1$). Although this requires the use of thermodynamic relations that hold only for elastic systems, we expect that this approximation could be reliable for not-strong values of inelasticity in collisions. Further comparisons with computer simulations are needed to confirm the above expectation. 
For molecular mixtures, the expression for the chemical potential of the impurities consistent with the approximation \eqref{4.12} is
\beqa
\label{4.13}
\frac{\mu_0}{T}&=&\ln(\lambda_0^2 n_0)+\ln n_0-\ln (1-\phi)\nonumber\\
& & +\frac{1}{4}\omega\Bigg[\frac{9\phi}{1-\phi}+\ln (1-\phi)\Bigg]
\nonumber\\
& &
-\frac{\omega^2}{8}\Bigg[\frac{\phi(1-10\phi)}{(1-\phi)^2}-\frac{8\phi}
{1-\phi}+\ln (1-\phi)\Bigg],\nonumber\\
\eeqa
where $\lambda_0(T)$ is the (constant) de Broglie's thermal wavelength.\cite{RG73} The use of Eq.\ \eqref{4.13} allows us to compute the derivative $(\partial \mu_0/\partial \phi)_{T,n_0}$ in Eq.\ \eqref{3.11}. 
}

As for elastic collisions,\cite{CC70} to get explicit expressions for the diffusion transport coefficients one considers the leading terms in a Sonine polynomial expansion of the unknowns $\boldsymbol{\mathcal{A}}_{0}$, $\boldsymbol{\mathcal{B}}_{0}$, and $\boldsymbol{\mathcal{C}}_{0}$. This task is carried out in Sec.\ \ref{sec5}.

\section{First-Sonine approximation to the diffusion transport coefficients at the stationary temperature}
\label{sec5}

The lowest order Sonine polynomial approximations for $\boldsymbol{\mathcal{A}}_{0}\left(\mathbf{V}\right)$, $\boldsymbol{\mathcal{B}}_{0}\left(\mathbf{V}\right)$, and $\boldsymbol{\mathcal{C}}_{0}\left(\mathbf{V}\right)$ are
\beq
\label{4.1}
\boldsymbol{\mathcal{A}}_{0}\left(\mathbf{V}\right) \to -f_{0,\text{M}}(\mathbf{V})\frac{\rho}{n_0 T_0^{(0)}} \mathbf{V}\; D_T,
\eeq
\beq
\label{4.2}
\boldsymbol{\mathcal{B}}_{0}\left(\mathbf{V}\right) \to -f_{0,\text{M}}(\mathbf{V})\frac{m_0^2}{\rho T_0^{(0)}} \mathbf{V}\; D_0,
\eeq
\beq
\label{4.3}
\boldsymbol{\mathcal{C}}_{0}\left(\mathbf{V}\right) \to -f_{0,\text{M}}(\mathbf{V})\frac{m_0}{n_0 T_0^{(0)}} \mathbf{V}\; D,
\eeq
where $T_0^{(0)}$ is the zeroth-order contribution to the partial temperature $T_0$ and the Maxwellian distribution $f_{0,\text{M}}(\mathbf{V})$ is given by Eq.\ \eqref{2.6.1} with the replacements $\mathbf{v}\to \mathbf{V}$ and $T_0\to T_0^{(0)}$. Moreover, in the lowest Sonine approximation, $\boldsymbol{\mathcal{A}}=\boldsymbol{\mathcal{C}}=\mathbf{0}$, and then  the impurities do not inherit any \RG{first-order} contribution coming from the granular gas. \vicente{This means that in the lowest Sonine approximation the integral equations \eqref{4.1} and \eqref{4.3} for  
$\boldsymbol{\mathcal{A}}_0$ and $\boldsymbol{\mathcal{C}}_0$, respectively, are not coupled to their corresponding counterparts $\boldsymbol{\mathcal{A}}$ and $\boldsymbol{\mathcal{C}}$ of the granular gas}. 

The transport coefficients $D_0$, $D$, and $D_T$ are determined by substitution of Eqs.\ \eqref{4.1}--\eqref{4.3} into the integral equations \eqref{3.6}--\eqref{3.8}, multiplication of them by $m_0 \mathbf{V}$ and integration over velocity. Some technical details on this calculation are displayed in the Appendix \ref{appA}. Here, we provide the final expressions for the coefficients $D_0$, $D$, and $D_T$ in the relevant state of a two-dimensional confined granular mixture with stationary temperature. In this case, $\zeta^{(0)}=\zeta_0^{(0)}=0$ and the differential equations \eqref{a3}, \eqref{a10}, and \eqref{a12} obeying the diffusion coefficients (in dimensionless form) become linear algebraic equations.

The expressions of the diffusion transport coefficients $D_0$, $D_T$, and $D$ in the steady state are given, respectively, by
\beq
\label{4.4}
D_0=\frac{\rho T}{m_0^2 \nu}\frac{\gamma_0}{\nu_D^*},
\eeq
\beq
\label{4.5}
D_T=\frac{n_0 T}{\rho \nu}\frac{X_0^*}{\nu_D^*+\frac{1}{2}{\Delta}^*\frac{\partial \zeta^*}{\partial {\Delta}^*}},
\eeq
\beq
\label{4.6}
D=\frac{n_0 T}{m_0 \nu}\frac{Y_0^*}{\nu_D^*}.
\eeq
In Eqs.\ \eqref{4.4}--\eqref{4.6}, the dimensionless quantities $\nu_D^*$, $X_0^*$, and $Y_0^*$ are given, respectively, by 
\beq
\label{4.7}
\nu_\text{D}^*=\frac{2\pi^{\frac{d-1}{2}}}{d\Gamma\left(\frac{d}{2}\right)}\chi_0 \left(\frac{\bar{\sigma}}{\sigma}\right)^{d-1} M \Bigg[\left(\frac{1+\theta}{\theta}\right)^{1/2}(1+\al_0)+\sqrt{\pi}\Delta_0^*\Bigg],
\eeq
\beqa
\label{4.8}
X_0^*&=&\gamma_0\Big(1-\frac{1}{2}
\widetilde{\Delta}^*\frac{\partial \ln \gamma_0}{\partial \widetilde{\Delta}^*}\Big)-\mu\Big(p^*-\frac{2^{d-1}}{\sqrt{2\pi}}\chi \phi \Delta^*\Big)\nonumber\\
& &
+2^d \left(\frac{\bar{\sigma}}{\sigma}\right)^{d} \phi \chi_0 M_0 \Bigg[\frac{1+\al_0}{2}+\frac{\Delta_0^*}{\sqrt{\pi}}
\left(\frac{\theta}{1+\theta}\right)^{1/2}\Bigg],
\nonumber\\
\eeqa
\beqa
\label{4.9}
Y_0^*&=&-\mu \left(p^*+\phi \frac{\partial p^*}{\partial \phi}\right)+ \phi M_0 \left(\frac{\partial \mu_0/T}{\partial \phi}\right)_T
\left(\frac{1+\theta}{\theta}\right)\nonumber\\
& & \times \Bigg[\frac{1+\al_0}{2}+\frac{2\Delta_0^*}{\sqrt{\pi}}
\left(\frac{\theta}{1+\theta}\right)^{1/2}\Bigg],
\eeqa
where $\gamma_0=T_0^{(0)}/T$, $\mu=m_0/m$ is the mass ratio, and we have introduced the shorthand notation
\beq
\label{4.10}
\widetilde{\Delta}^*\frac{\partial}{\partial \widetilde{\Delta}^*}\equiv \Delta^*\frac{\partial}{\partial \Delta^*}+\Delta_{0}^*\frac{\partial}{\partial \Delta_{0}^*}.
\eeq
Note that in the particular case $\Delta^*=\Delta_{0}^*$, only one of the two terms of the identity \eqref{4.10} must be considered. The evaluation of the derivative $\widetilde{\Delta}^*\partial_{\widetilde{\Delta}^*}$ appearing in Eq.\ \eqref{4.8} is performed in the Appendix \ref{appA}.

\begin{figure}
\begin{center}
\begin{tabular}{lr}
\resizebox{6.5cm}{!}{\includegraphics{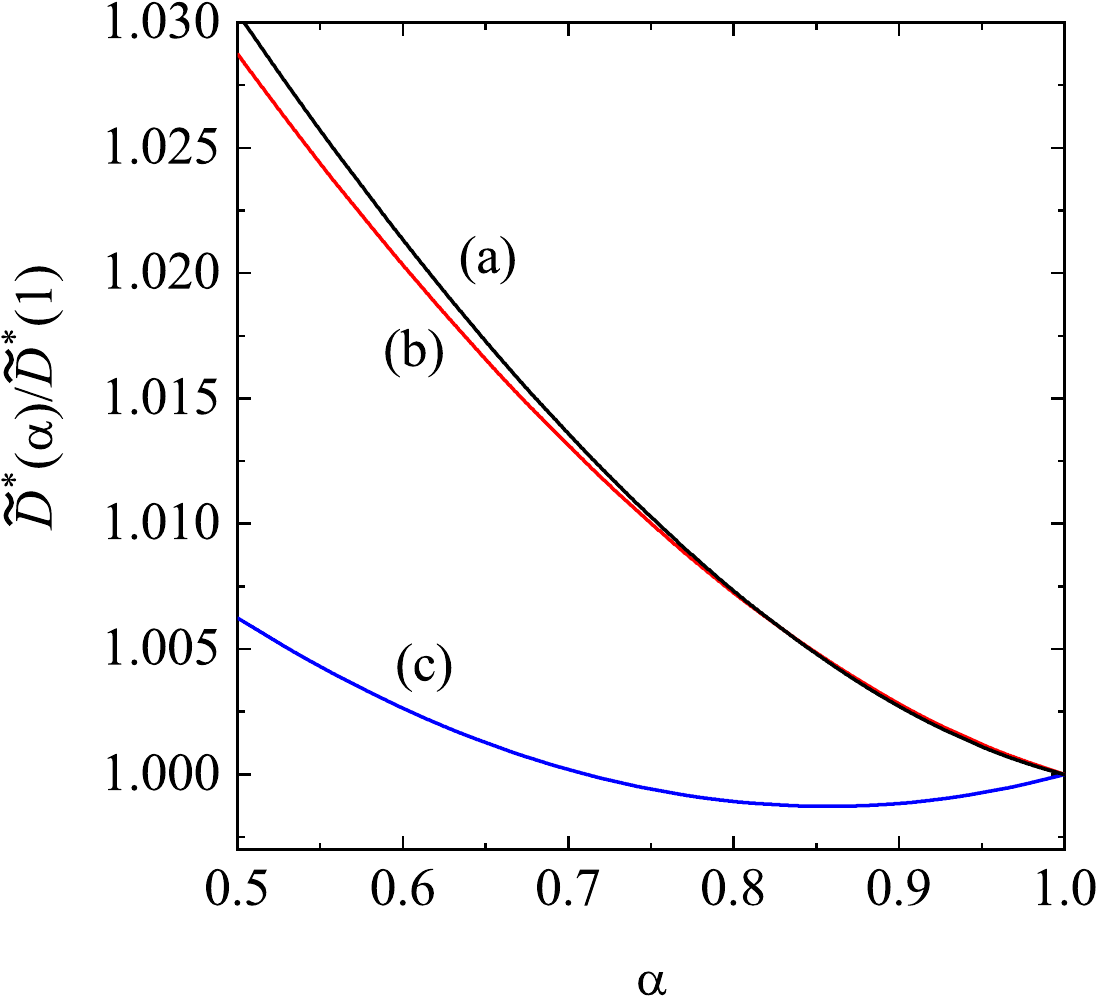}}
\end{tabular}
\begin{tabular}{lr}
\resizebox{6.5cm}{!}{\includegraphics{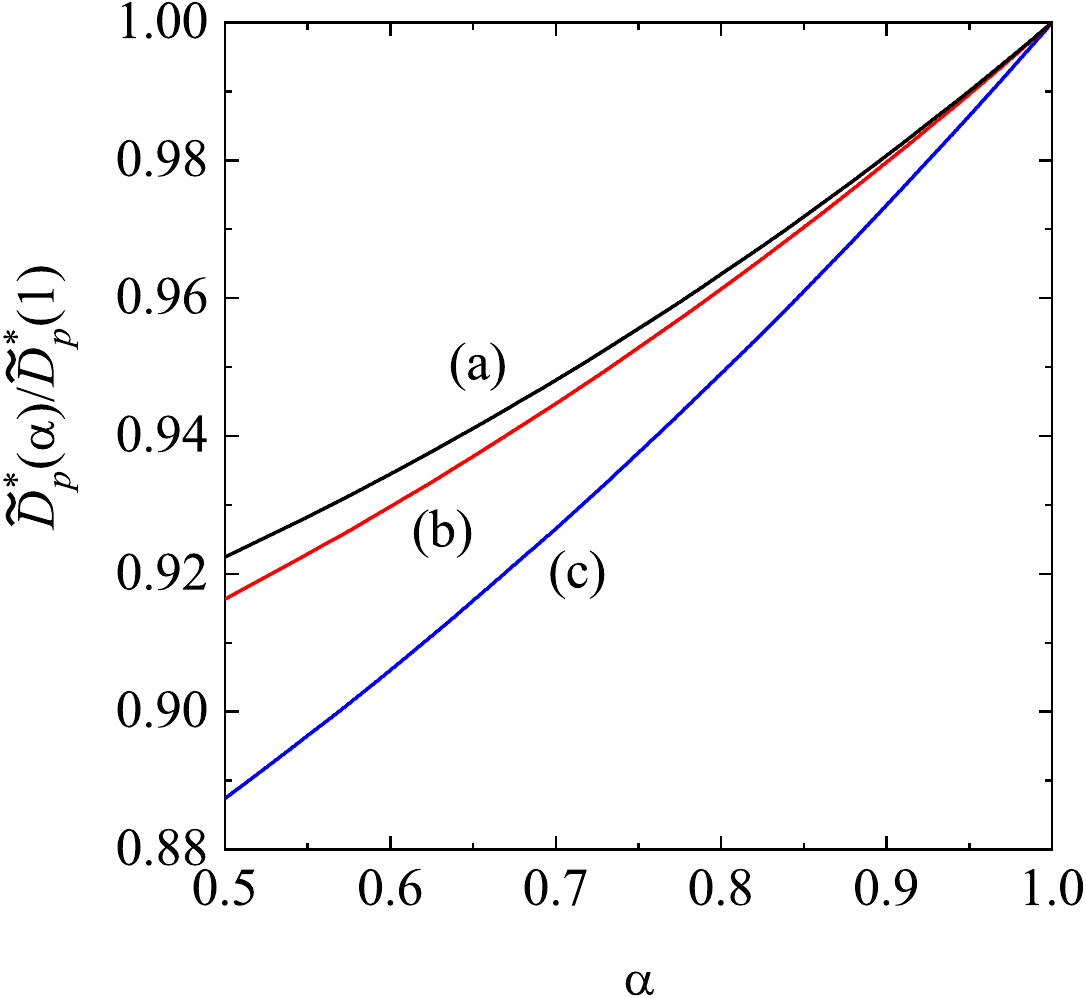}}
\end{tabular}
\begin{tabular}{lr}
\resizebox{6.5cm}{!}{\includegraphics{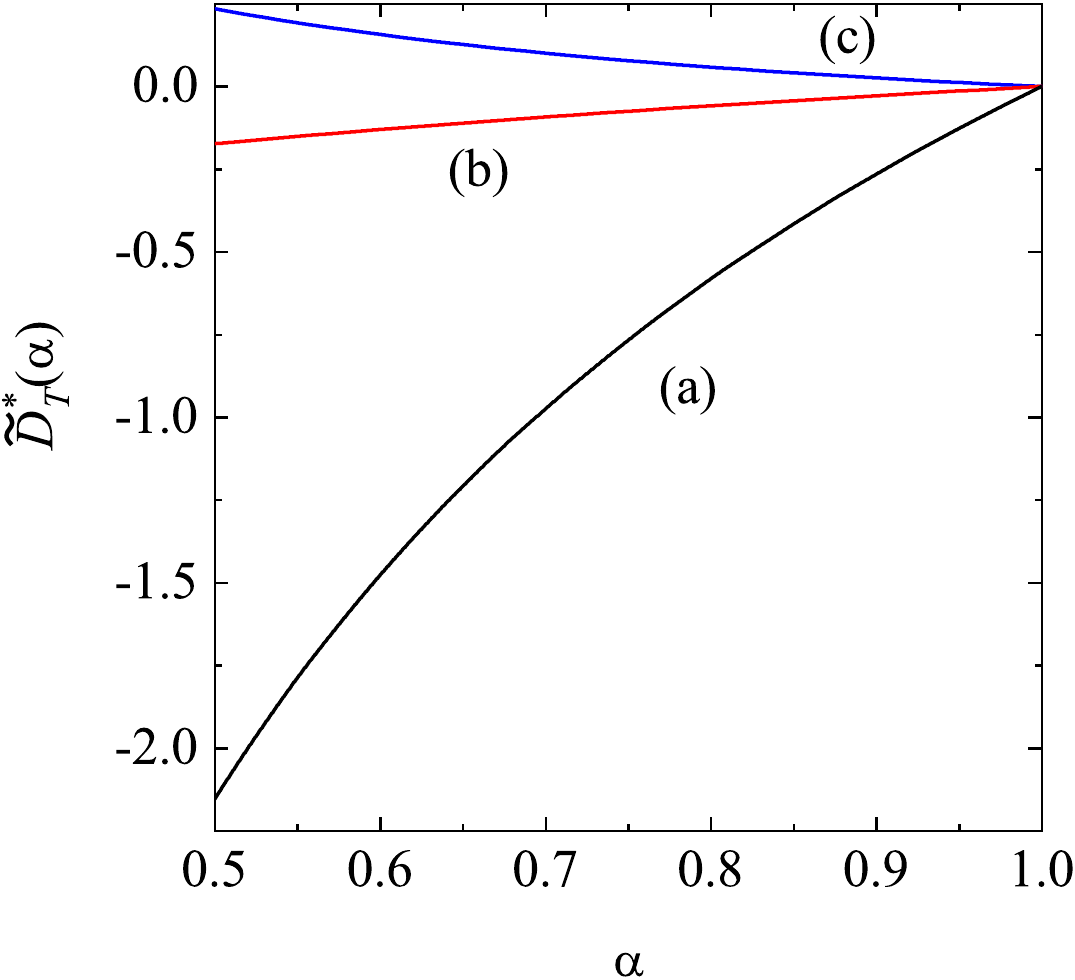}}
\end{tabular}
\end{center}
\caption{Plot of the (dimensionless) diffusion coefficients $\widetilde{D}^*(\alpha)/\widetilde{D}^*(1)$, $\widetilde{D}_p^*(\alpha)/\widetilde{D}_p^*(1)$, and $\widetilde{D}_T^*(\al)$ vs the (common) coefficient of restitution $\al=\al_0$ for $d=2$, $\phi=0$, and three different binary mixtures: $\sigma_0/\sigma=2$, $m_0/m=4$ (a); $\sigma_0/\sigma=2$, $m_0/m=2$ (b); and $\sigma_0/\sigma=0.2$, $m_0/m=0.8$ (c). Here, $\Delta^*=\Delta_0^*$ and $\widetilde{D}^*(1)$
and $\widetilde{D}_p^*(1)$ refer to the values of the diffusion coefficients for elastic collisions ($\al=1$).
\label{fig1}}
\end{figure}

Before considering some illustrative systems to assess the dependence of the diffusion coefficients on the parameter space of the system, it is quite instructive to pay attention to some limiting cases.

\subsection{Mechanically equivalent particles}

For mechanically equivalent particles ($m=m_0$, $\sigma=\sigma_0$, $\al=\al_0$, and $\Delta=\Delta_0$), it is easy to see that $X_0^*=0$ and $Y_0^*=-1$. In this case, $D_T=0$ and $D=-(n_0/n)D_0$, as expected. The mass flux $\mathbf{j}_0$ obeys the constitutive equation
\beq
\label{4.14}
\mathbf{j}_0=-m D_0 \nabla x_0,
\eeq
where 
\beq
\label{4.14.1}
D_0=\frac{d\Gamma\left(\frac{d}{2}\right)}{2\pi^{(d-1)/2}}
\frac{\sigma^{1-d}}{\chi}
\sqrt{\frac{T}{m}}
\left(1+\al+\sqrt{\frac{\pi}{2}}\Delta^*\right)^{-1}
\eeq
is the self-diffusion coefficient.

\subsection{Low-density limit}

Let us consider now the low-density limit ($\phi\to 0$). In this case, to compare with the results derived in Ref.\ \onlinecite{GBS21} for general concentration (or mole fraction) $x_0\equiv n_0/n$, it is convenient to express the constitutive equation \eqref{0.1} in terms of the spatial gradients of $x_0$, $p=nT$ and $T$. In this representation,
\beq
\label{4.15}
\mathbf{j}_0^{(1)}=-\frac{m m_0 n}{\rho}\widetilde{D} \nabla x_0-\frac{\rho}{p} \widetilde{D}_p \nabla p-\frac{\rho}{T} \widetilde{D}_T \nabla T,
\eeq
where the relationship between the diffusion coefficients $\left\{\widetilde{D}, \widetilde{D}_p, \widetilde{D}_T\right\}$ and $\left\{D_0, D, D_T\right\}$ is
\beq
\label{4.16}
D_0=\mu^{-1}\widetilde{D}, \quad D=\frac{\rho}{m_0}\widetilde{D}_p-x_0 \widetilde{D}, \quad D_T=\widetilde{D}_p+\widetilde{D}_T.
\eeq

In the dilute limit ($\phi\to 0$), Eqs.\ \eqref{4.4}--\eqref{4.6} for the coefficients $D_0$, $D_T$, and $D$ reduce to
\beq
\label{4.17}
D_0=\frac{\rho T}{m_0^2 \nu}\frac{\gamma_0}{\nu_D^*}, \quad D=-\frac{n_0 T}{m_0 \nu}\frac{\mu}{\nu_D^*},
\eeq
\beq
\label{4.18}
D_T=\frac{n_0 T}{\rho \nu}
\frac{\gamma_0\Big(1-\frac{1}{2}
\widetilde{\Delta}^*\frac{\partial \ln \gamma_0}{\partial \widetilde{\Delta}^*}\Big)-\mu}{\nu_D^*+\frac{1}{2}{\Delta}^*\frac{\partial \zeta^*}{\partial {\Delta}^*}},
\eeq
where $\nu_D^*$ is given by Eq.\ \eqref{4.7} with $\chi_0=1$. Substitution of Eqs.\ \eqref{4.17} and \eqref{4.18} into the relationships \eqref{4.16} yield
\beq
\label{4.19}
\widetilde{D}=\frac{\rho T}{m m_0 \nu}\frac{\gamma_0}{\nu_D^*},
\eeq
\beq
\label{4.20}
\widetilde{D}_p=x_0\frac{n T}{\rho \nu}\frac{\gamma_0-\mu}{\nu_D^*},
\eeq
\beq
\label{4.21}
\widetilde{D}_T=-x_0\frac{n T}{\rho \nu}\frac{
\widetilde{\Delta}^*\frac{\partial \gamma_0}{\partial \widetilde{\Delta}^*}+{\Delta}^*\frac{\partial \zeta^*}{\partial {\Delta}^*}\widetilde{D}_p^*}{2\nu_D^*+{\Delta}^*\frac{\partial \zeta^*}{\partial {\Delta}^*}},
\eeq
where $\widetilde{D}_p^*=(\rho \nu/x_0 nT)\widetilde{D}_p$. Equations \eqref{4.19}--\eqref{4.21} are consistent with those obtained for dilute granular mixtures in the tracer limit ($x_0\to 0$). \cite{GBS21}

The dependence of the (dimensionless) diffusion transport coefficients $\widetilde{D}^*(\alpha)/\widetilde{D}^*(1)$, $\widetilde{D}_p^*(\alpha)/\widetilde{D}_p^*(1)$, and $\widetilde{D}_T^*(\al)$ on a (common) coefficient of restitution $\al=\al_0$ is plotted in Fig.\ \ref{fig1} for three different two-dimensional ($d=2$) mixtures.
\RS{Note that as the plotted quantities are dimensionless, they cannot depend on dimensional values like $\Delta$, but only in their dimensionless form $\Delta^*$. As the system is in the HSS, the latter is given by Eq.~\eqref{1.17} and, therefore, all dimensionless quantities can be plotted simply as a function of the restitution coefficient. Subsequent figures will adopt the same convention.
}
In Fig.\ \ref{fig1}, $\Delta^*=\Delta_0^*$, $\widetilde{D}^*=(m m_0 \nu/\rho T)\widetilde{D}$, and $\widetilde{D}_T^*=(\rho\nu/x_0 n T)\widetilde{D}_T$. Moreover, $\widetilde{D}^*(1)$ and $\widetilde{D}_p^*(1)$ refer to the values of the diffusion coefficients for elastic collisions ($\al=1$). The (dimensionless) thermal diffusion coefficient $\widetilde{D}_T^*$ has not been reduced with its elastic value because this coefficient vanishes when $\al=1$ for \emph{dilute} gases in the first Sonine approximation. \cite{CC70,KCL87} As already noted in previous works, \cite{GBS21} Fig.\ \ref{fig1} shows that
the impact of inelasticity on the diffusion coefficients is in general smaller than the one found in the conventional IHS model. \cite{GMD06} Regarding the tracer diffusion coefficient $\widetilde{D}^*$, we see this coefficient increases with respect to its elastic value as the inelasticity increases, except when the impurity is smaller and/or lighter than the granular gas particles; there exists a small region around $\al=1$ where the scaled coefficient $\widetilde{D}^*(\al)/\widetilde{D}^*(1)$ exhibits a non-monotonic dependence on inelasticity. In the case of the (scaled) coefficient $\widetilde{D}_p^*(\al)/\widetilde{D}_p^*(1)$ we observe that it decreases with decreasing $\al$ (regardless of the value of the diameter and/or mass ratio  while the thermal diffusion coefficient $\widetilde{D}_T^*(\al)$ can be positive or negative depending on the system considered. This behavior agrees qualitatively to what happens in the IHS model. \cite{GBS21} As will show later, the signature of the coefficient $\widetilde{D}_T^*(\al)$ is important in segregation problems induced by a thermal gradient. \cite{JY02,BRM05,BRM06,SGNT06,G06,G08a,BEGS08,G11,GG23}

\begin{figure}
\begin{center}
\begin{tabular}{lr}
\resizebox{6.5cm}{!}{\includegraphics{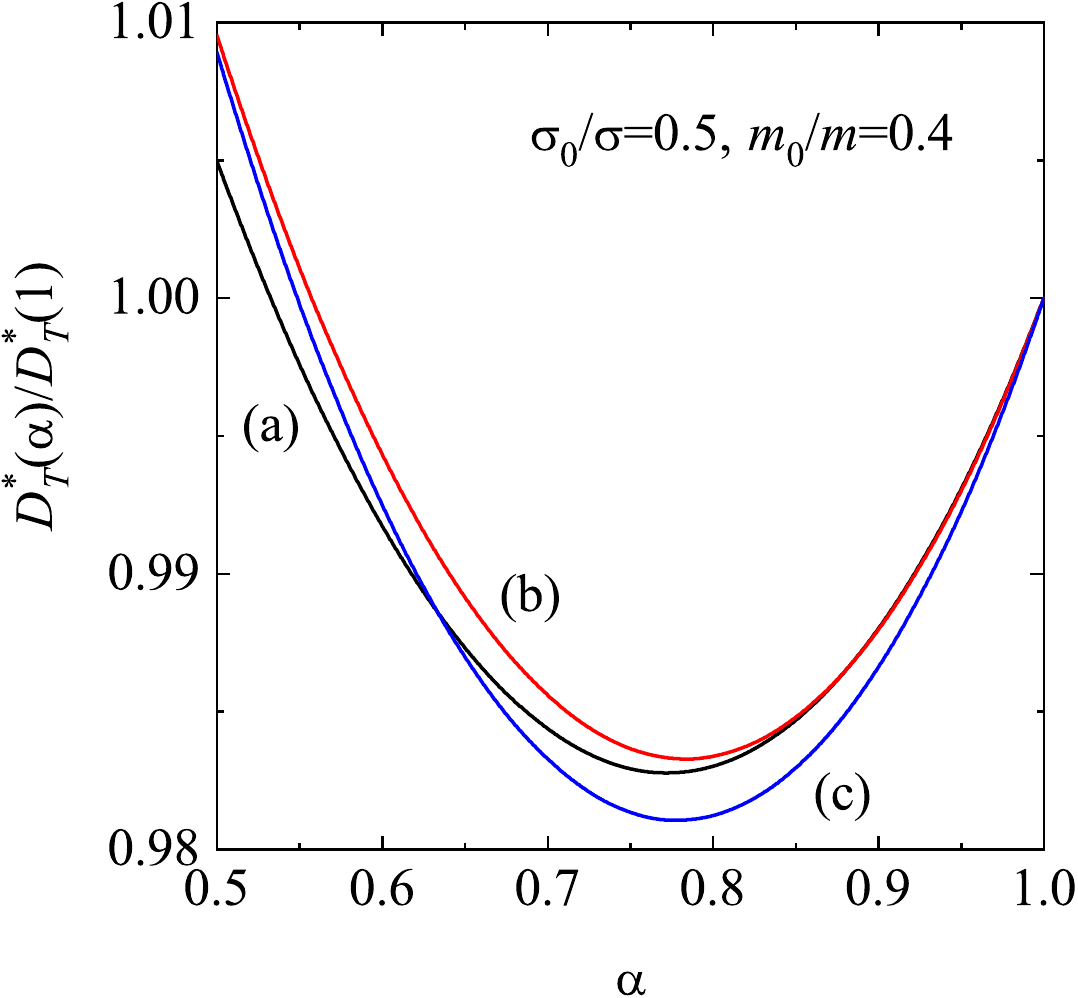}}
\end{tabular}
\begin{tabular}{lr}
\resizebox{6.5cm}{!}{\includegraphics{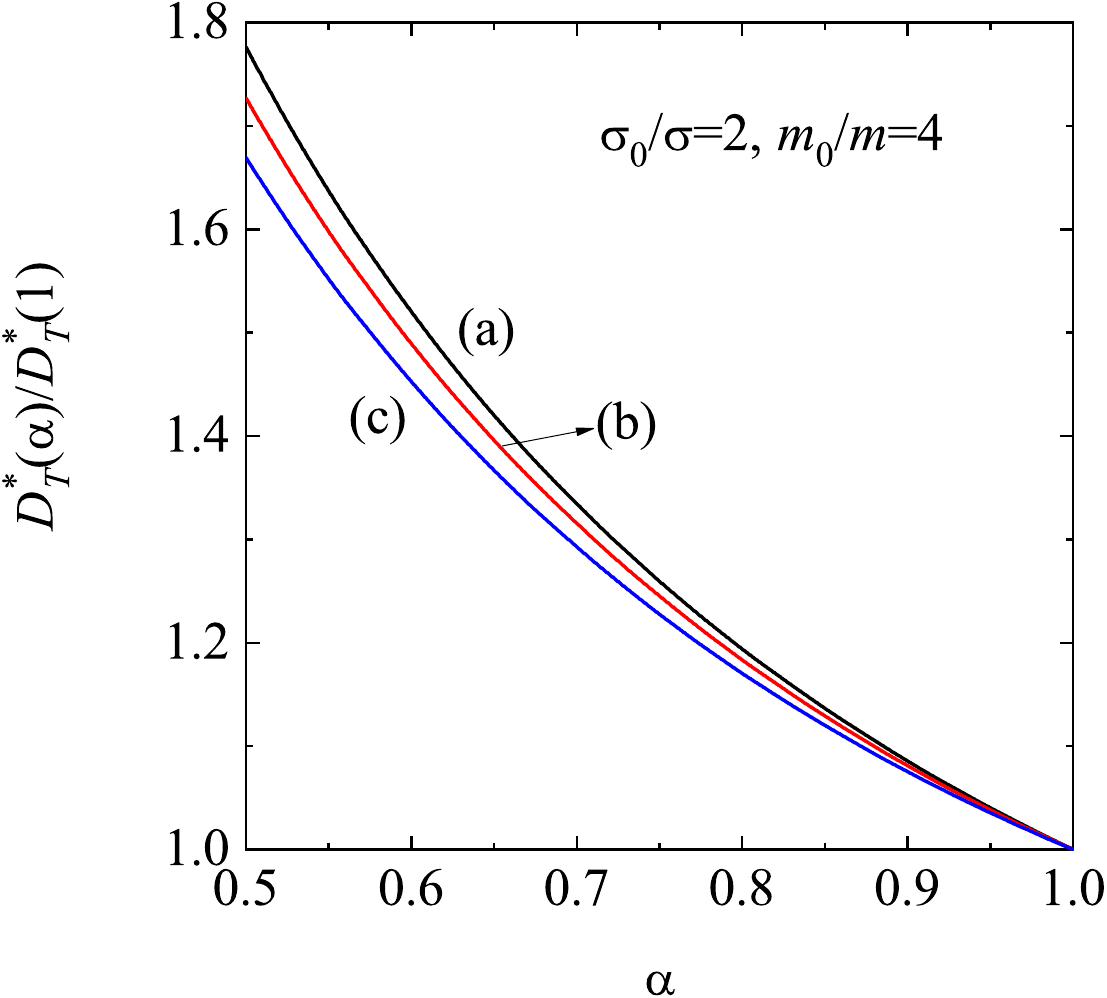}}
\end{tabular}
\end{center}
\caption{Plot of the (dimensionless) transport diffusion coefficient $D_T^*(\alpha)/D_T^*(1)$
vs the (common) coefficient of restitution $\al=\al_0$ for $d=2$, two different binary mixtures
($\sigma_0/\sigma=2$, $m_0/m=4$ and $\sigma_0/\sigma=2$, $m_0/m=0.4$) and three different values of
the solid volume fraction $\phi$: $\phi=0$ (a), $\phi=0.1$ (b), and $\phi=0.2$ (c). Here, $\Delta^*=\Delta_0^*$ and $D_T^*(1)$
refers to the value of $D_T^*$ for elastic collisions ($\al=1$).
\label{fig2}}
\end{figure}

\begin{figure}
\begin{center}
\begin{tabular}{lr}
\resizebox{6.5cm}{!}{\includegraphics{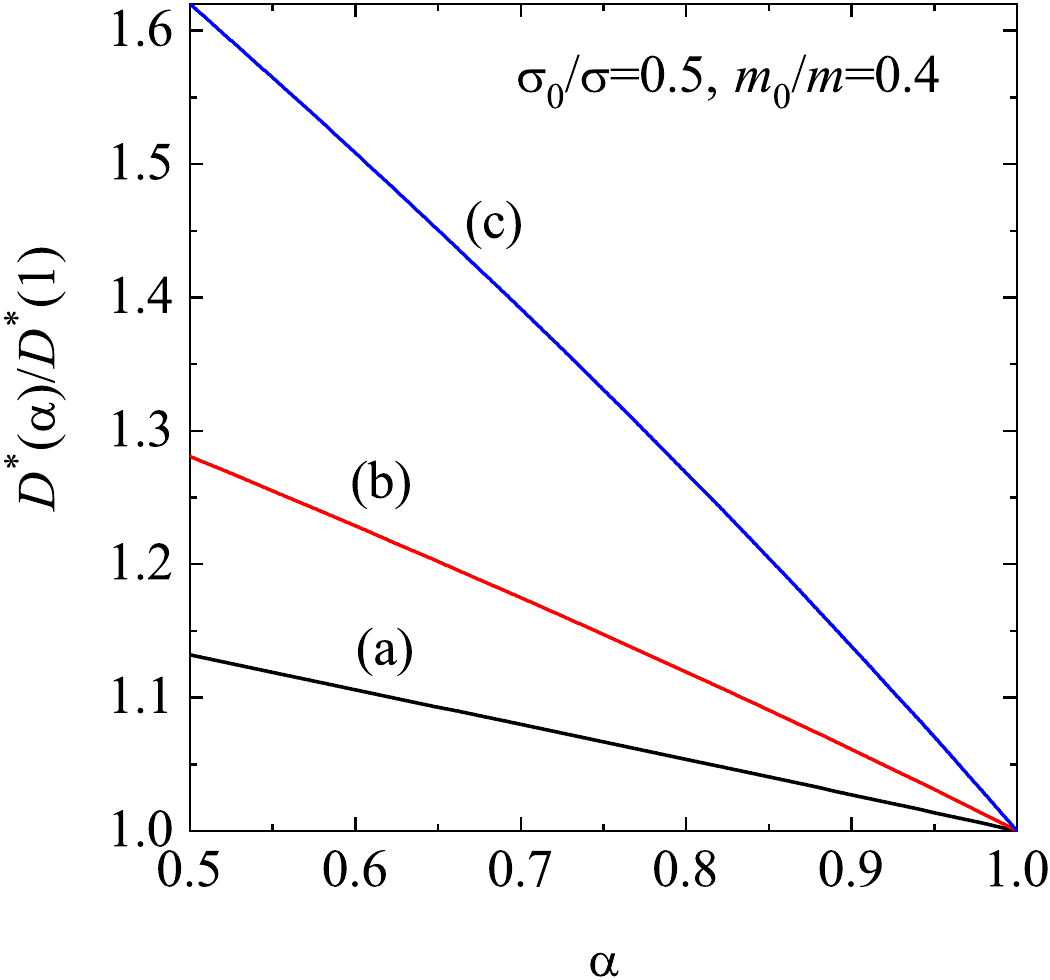}}
\end{tabular}
\begin{tabular}{lr}
\resizebox{6.5cm}{!}{\includegraphics{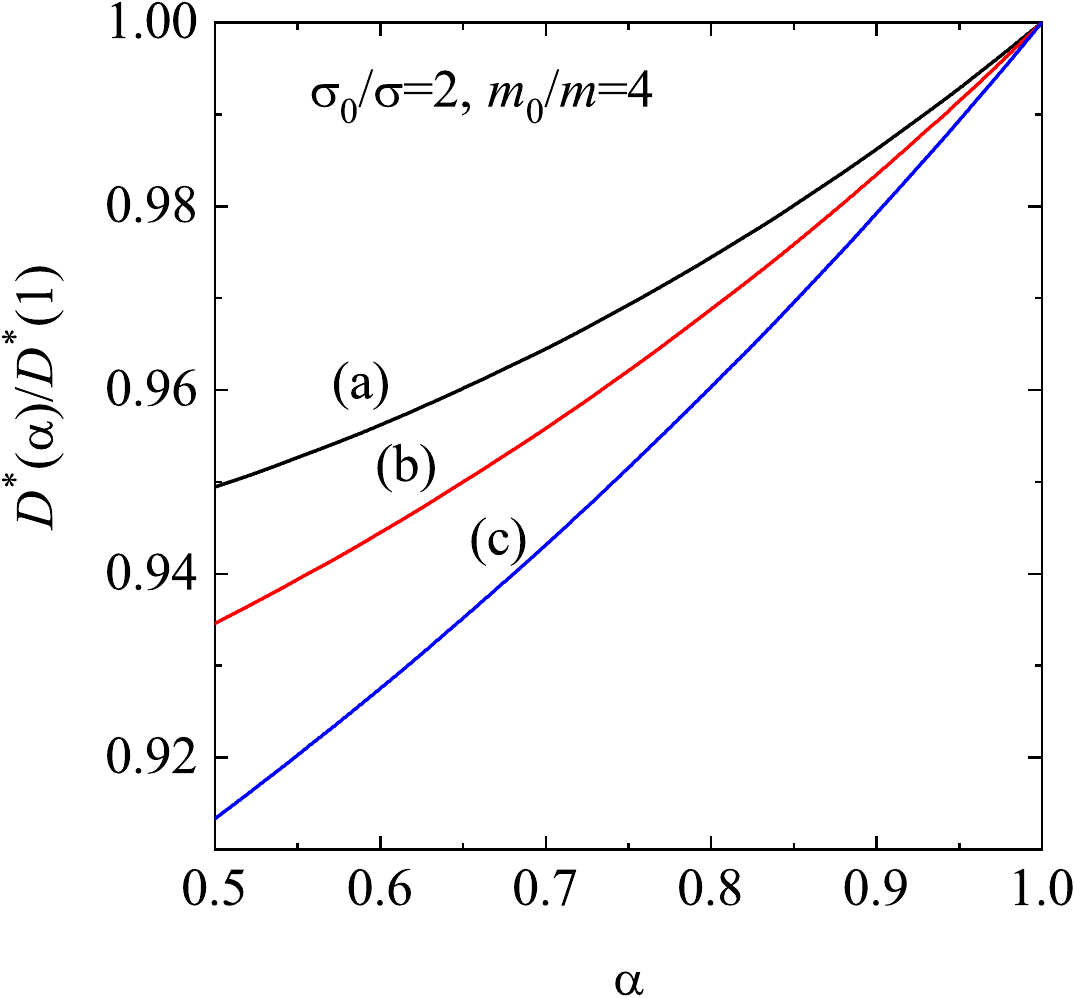}}
\end{tabular}
\end{center}
\caption{Plot of the (dimensionless) transport diffusion coefficient $D^*(\alpha)/D^*(1)$ vs the
(common) coefficient of restitution $\al=\al_0$ for $d=2$, two different binary mixtures ($\sigma_0/\sigma=2$,
$m_0/m=4$ and $\sigma_0/\sigma=2$, $m_0/m=0.4$) and three different values of the solid volume fraction $\phi$: $\phi=0$ (a),
$\phi=0.1$ (b), and $\phi=0.2$ (c). Here, $\Delta^*=\Delta_0^*$ and $D^*(1)$ refers to the value of $D^*$
for elastic collisions ($\al=1$).
\label{fig3}}
\end{figure}

\subsection{Some illustrative systems for moderate densities}

Now, we want to assess the dependence of the (scaled) diffusion coefficients on the parameter space of the system for a two-dimensional system. The parameter space is constituted by the diameter $\sigma_0/\sigma$ and mass $m_0/m$ ratios, the coefficients of restitution $\al$ and $\al_0$, the solid volume fraction $\phi$, and the (dimensionless) parameters associated with the energy injection at collisions  $\Delta^*$ and $\Delta_0^*=\lambda \Delta^*$ ($\lambda \geq 0$). The parameter $\lambda$ is therefore a measure in the contrast of energy injection at tracer-gas collisions compared to gas-gas collisions. Since we are mainly interested on the influence of both inelasticity and density on mass transport, we first analyze the usual case for binary mixtures, that is,
when the components differ only in their masses and diameters. Hence, we assume here that $\al=\al_0$ and $\lambda=1$. In addition, because in the steady state $\Delta^*$ is only a function of $\al$ [see Eq.\ \eqref{1.17}], then the parameter space is reduced to three parameters $\left\{\sigma_0/\sigma, m_0/m, \al\right\}$.

According to Eq.\ \eqref{4.4}, the (dimensionless) tracer diffusion coefficient $D_0^*=(m_0^2\nu/\rho T)D_0$ is given by  
\beq
\label{4.22}
D_0^*=\frac{\gamma_0}{\chi_0 \nu_D^*}.
\eeq
Since the temperature ratio $\gamma_0$ is obtained from the condition $\zeta_0=0$, then it is independent of the density $\phi$. Consequently, according to Eq.\ \eqref{4.7}, the scaled coefficient $D_0^*(\al)/D_0^*(1)$ is independent of the density and hence, its dependence on $\al$ is the same as that of a dilute gas (see Fig.\ \ref{fig1}). For this reason, we focus our attention now in the scaled coefficients $D_T^*(\al)/D_T^*(1)$ and $D^*(\al)/D^*(1)$, where $D_T^*=(\rho \nu/n_0 T)D_T$ and $D^*=(m_0 \nu/n_0 T)D$. As said before, $D_T^*(1)$ and $D^*(1)$ refer to the values of $D_T^*$ and $D^*$, respectively, for elastic collisions ($\al=1$). According to Eq.\ \eqref{4.8}, note that for dense gases $X_0^*\neq 0$ and so, $D_T^*(1)\neq 0$ when $\phi\neq 0$.

Figure \ref{fig2} shows the dependence of the ratio $D_T^*(\al)/D_T^*(1)$ on $\al$ for three different values of the density $\phi$. Two different mixtures are considered. We observe first that while this scaled coefficient exhibits a non-monotonic dependence on inelasticity when both mass and diameter ratios are smaller than 1, the ratio $D_T^*(\al)/D_T^*(1)$ increases with increasing inelasticity when the impurity is heavier and/or larger than the particles of the granular gas. This behavior contrasts with the one found in the conventional IHS (see the lines of the first Sonine solution of Fig.\ 3 in Ref.\ \onlinecite{GV09}). With respect to the influence of density, at a given value of $\al$, we see in general a weak effect of $\phi$ on the (scaled) thermal diffusion coefficient. To complement this figure, the (scaled) coefficient $D^*(\al)/D^*(1)$ is plotted in Fig.\ \ref{fig3} as a function of $\al$ for the same systems as those considered in Fig.\ \ref{fig2}. In contrast to $D_T^*$, stronger influence of density on $D^*(\al)/D^*(1)$ is observed. At a given value of $\al$, while this scaled coefficient increases as the granular gas becomes denser when $m_0>m$ and $\sigma_0>\sigma$, the opposite happens when the impurity is lighter and/or smaller than the particles of the granular gas. As in the case of $D_T^*$, a comparison with the results obtained in the IHS model (see the lines of the first Sonine solution of Fig.\ 2 in Ref.\ \onlinecite{GV09}) shows again a completely different qualitative behavior. As a general conclusion, our results clearly show that the impact of inelasticity on mass transport for a confined system (modeled via the $\Delta$ model) is less significant than the one observed in freely cooling systems. This is the expected result based on the previous work \cite{GBS21} on dilute granular mixtures.

\section{Comparison with computer simulations}
\label{sec6}

The theoretical results derived for the diffusion transport coefficients are based on a relatively crude approximation: they have been obtained by considering the leading term in a Sonine polynomial expansion. Thus, it seems convenient to assess the accuracy of the above theoretical results by comparing them with computer simulations. Here, we compare the expression \eqref{4.4} for the tracer diffusion coefficient $D_0$ with both numerical results obtained from the DSMC method~\cite{B94} and from MD simulations.~\cite{TS05,AT17,FS23}  In this case, we consider the diffusion of tracer particles in a granular gas of mechanically different particles in the HSS. In this situation, $\nabla \mathbf{U}=\mathbf{0}$, $\nabla n=\nabla T=0$ and so, the constitutive equation \eqref{3.2} can be written as
\beq
\label{5.1}
\mathbf{j}_0^{(1)}=-\frac{m_0^2 n}{\rho}D_0 \nabla x_0,
\eeq
where we recall that $x_0=n_0/n$. Then, the balance equation \eqref{2.4} for the concentration becomes    
\beq
\label{5.2}
\frac{\partial x_0}{\partial t}=\frac{m_0 D_0}{\rho}\nabla^2 x_0.
\eeq
Equation \eqref{5.2} is the standard diffusion equation with the time-independent diffusion coefficient $D_\text{tracer}=m_0 D_0/\rho$. In simulations, we can obtain  the Mean Square Displacement (MSD) of the tracer particles after a time interval $t$ as 
\beq
\label{5.2.1}
\text{MSD}(t)=\langle |\mathbf{r}(t)-\mathbf{r}(0)|^2\rangle,
\eeq
where $|\mathbf{r}(t)-\mathbf{r}(0)|$ is the distance travelled by the tracer particles from $t=0$ until the instant $t$.
From it,  the Einstein relation gives
\beq
\label{5.2b}
D_\text{tracer}= \frac{1}{2d}\frac{d\, \text{MSD}(t)}{d t}.
\eeq
Equation \eqref{5.2b} finally permits us to obtain the diffusion coefficient $D_0$.

The DSMC method is a direct particle numerical method to find solutions to the Boltzmann \cite{B94} and Enskog \cite{MS96,MS97} kinetic equations, without the need of considering uncontrolled approximations such as the truncation of a Sonine polynomial expansion. Thus, in the limit of an infinitely large number of particles and small time step, the DSMC method provides the ``exact'' solution of the kinetic equation. It is therefore an excellent method to assess the reliability of the theoretical expression \eqref{4.4} for $D_0$ obtained from the first Sonine approximation. In practice, we consider $N=4000$ particles, one of which is the tracer particle, simulated in homogeneous conditions. This means that no spatial grid is constructed during the simulations and particles can collide with all of them. Collisions, which are sampled statistically, are performed following the collision rules \eqref{1.1}--\eqref{1.2} and \eqref{2.2.1bis}--\eqref{2.2.2bis}. 
Since the granular gas is in the HSS, the influence of the gas density $n\sigma^d$ on diffusion in the DSMC method enters only in fixing the collision rate. Thus, as in the Enskog theory, the density dependence in the DSMC method appears through the pair correlation factors at contact $\chi$ and $\chi_0$ given by Eqs.\ \eqref{4.11} and \eqref{4.12}, respectively. \RS{For the normalized transport coefficients $D_0^*(\alpha)/D_0^*(1)$, the Enskog theory predicts that the result is independent of density as the collision frequencies cancel. This allows us to use a single value for the density, with a value that is immaterial as it only fixes the time scale.}
Finally, the time step is set equal to one fiftieth of the mean collision time in the gas.

MD simulations can be considered as an exact numerical solution of the particle dynamics, following the collision rules \eqref{1.1}--\eqref{1.2} and \eqref{2.2.1bis}--\eqref{2.2.2bis}. As such, it makes no assumptions of molecular chaos, homogeneity, or normal solutions. Its results allow us to put strong tests on the hypothesis behind the kinetic description made in this article. For MD, we consider $N=4000$ particles, one of them being the tracer, which are simulated using the event-driven method\cite{TS05,AT17,FS23} in a square box with periodic boundary conditions, and a box size adjusted to give the desired particle density.
Both in DSMC and MD, large simulations are done to obtain convergence of the MSD.

\begin{figure}
    \centering
    \includegraphics[width=1\linewidth]{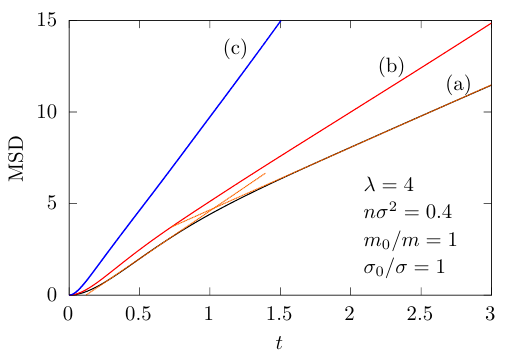}
    \caption{Mean-square displacements for tracer particles obtained in MD. The simulation parameters are $n\sigma^2=0.4$, $\lambda\equiv\Delta_0/\Delta=4$, and $\sigma_0/\sigma=m_0/m=1$, with a common coefficient of restitution $\alpha=\al_0=0.2$ (a), 0.5 (b), and 0.8 (c). 
    \RS{The two linear diffusive regimes for case (a) are shown in dotted orange lines, extrapolated beyond the diffusive regime to clearly show the different slopes.}
    \label{fig.MSD}}
\end{figure}

\subsection{Regimes in the mean-square displacement} \label{sec.regimesMSD}

Figure~\ref{fig.MSD} shows typical MSD curves obtained with MD simulations. It is evident a first ballistic regime \RS{associated to the free motion between collisions}, lasting for a time of the order of the mean collision time. At long times, a diffusive regime appears, which allows us to obtain  the diffusion coefficient using Einstein's relation~\eqref{5.2b}.
In cases of high density and small coefficients of restitution, three instead of two regimes appear (see, for example, the curves (a) and (b) in Fig.~\ref{fig.MSD}). First, there is the usual ballistic regime. It is followed by an intermediate diffusive regime with a diffusion coefficient $D_\text{transient}$ and, later, a second and definitive diffusive regime is established with a smaller diffusion coefficient $D_\text{tracer}$. The measurements presented in this paper are obtained in the last regime.

In the case of the DSMC method, only the ballistic and diffusion regimes are obtained (not shown). This implies that the origin of this anomalous behavior in MD can only be due to correlations neglected in the kinetic description. Specifically, as in Fig.~\ref{fig.MSD}, the value of $\lambda=4$ is chosen and hence the tracer-particle collision is characterized by a large value of $\Delta$. Therefore, the tracer particle acts as a local energy source, which is lately dissipated in the granular gas. As a result, the granular gas is driven into a \emph{inhomogeneous} state, hotter and more dilute near the tracer. This changes the local environment for the tracer particle, which can diffuse faster in this cloud, resulting in this larger value of $D_\text{transient}$. At longer times, the tracer itself must move the cloud, and the diffusion of the dressed object is described by the long term diffusion coefficient $D_\text{tracer}$.

\RS{Although the mechanism and the regimes found are different, the presence of various regimes in the MSD has been previously observed in MD simulations of freely cooling viscoelastic grains.~\cite{Bodrova2012}}
The analysis of this dynamical process \RS{and the relation with other systems presenting similar dynamics} is beyond the scope of this manuscript and is left to future work. Nevertheless, we would like to note that, despite its complexity, the numerical values obtained in MD for $D_\text{tracer}$ in general have a good quantitative agreement (except for a rather strong inelasticity) with the simple kinetic-theoretical prediction of $D_0$ given by the expression \eqref{4.4} (see especially figures ~\ref{fig.simcaseI} and \ref{fig.simcaseII} in the following section).

\subsection{Diffusion coefficients}

\begin{figure}
    \centering
    \includegraphics[width=1\linewidth]{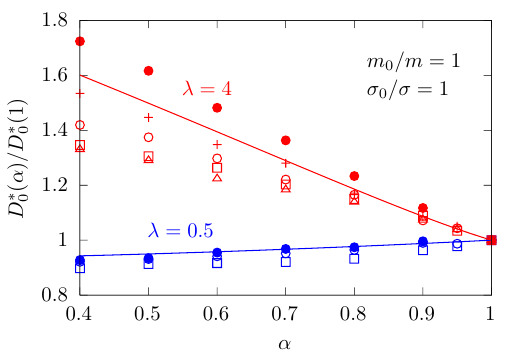}
    \caption{Comparison with simulations: case I. Normalized diffusion coefficient $D_0^*(\alpha)/D_0^*(1)$ as a function of the (common) coefficient of restitution $\alpha=\al_0$ for a tracer particle of equal mass and diameter of the particles of the granular gas, but with $\Delta_0=\lambda \Delta$, with $\lambda=0.5$ (blue) and $4.0$ (red). Theory in solid lines, DSMC results with solid circles, and MD simulations with open symbols ($n\sigma^2=0.05$, crosses; $n\sigma^2=0.1$, circles; $n\sigma^2=0.2$ triangles; and $n\sigma^2=0.4$ squares). 
    \label{fig.simcaseI}}
\end{figure}

We consider three different cases for study. In case I, the tracer particle has equal mass, diameter, and coefficient of restitution as the particles of the granular gas; it differs in the energy injection at collisions, with $\Delta_0=\lambda \Delta$. Figure~\ref{fig.simcaseI} presents the diffusion coefficient normalized to the elastic case $D_0^*(\alpha)/D_0^*(1)$ as a function of $\alpha$ for $\lambda=0.5$ and 4.
The theory predicts that the result is independent of density, result that is trivially obtained in DSMC because by construction of the method, the collision frequency scales out when the normalized diffusion coefficient is plotted. As a consequence, only one density is used for DSMC. In the case of MD simulations, several densities are used and there is an appreciable dependence with this parameter. We observe that the normalized diffusion coefficient decreases with increasing density, especially for the case of large contrast ($\lambda=4$). The DSMC results show good agreement with theory, with deviations increasing for more inelastic conditions. This is expected as the theoretical prediction has been obtained considering up to the first Sonine approximation and DSMC results can be considered an ``exact'' solution of the kinetic equation. As expected from previous works \cite{GM04,GV09} for tracer diffusion in the IHS model, going to larger inelasticities would require using more higher order approximations to achieve more accurate predictions. Regarding the MD simulations, the agreement is also good, except for the large contrast case, with large inelasticities and high densities. The deviations in these cases (which go in the opposite direction of the DSMC results) are essentially due to correlations not accounted for in the Enskog kinetic theory description. These include velocity correlations 
between the particles which are about to collide (breakdown of the molecular chaos hypothesis)
and the spatial inhomogeneities described in Sec.~\ref{sec.regimesMSD}. The effect of both on diffusion increases with the contrast parameter $\lambda$, inelasticities, and density.

\begin{figure}
    \centering
    \includegraphics[width=1\linewidth]{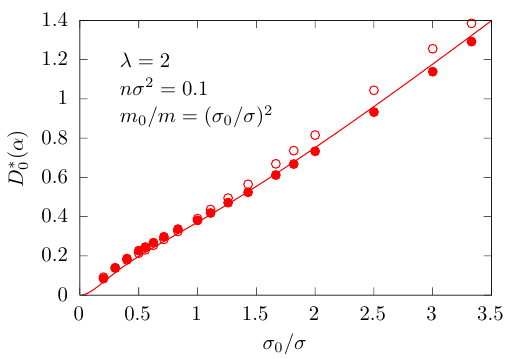}
    \caption{Comparison with simulations: case II. Diffusion coefficient $D_0^*(\alpha)$ as a function of the diameter ratio $\sigma_0/\sigma$ for a tracer particle with the same mass density as the particles of the granular gas [$m_0/m=(\sigma_0/\sigma)^2$]. The (common) coefficient of restitution is $\alpha=\al_0=0.8$, the contrast parameter  $\lambda\equiv\Delta_0/\Delta=2$, and the (reduced) density $n\sigma^2=0.1$. Theory in solid lines, MD simulations with open circles, and DSMC results with solid circles.}
    \label{fig.simcaseII}
\end{figure}

The case II considers impurities of different mass and radius compared to the particles of the granular gas, but sill having the same mass density (i.e., $m_0/\sigma_0^2=m/\sigma^2$). In Fig.~\ref{fig.simcaseII}, we present the diffusion coefficient $D_0^*(\alpha)$ as a function of the diameter ratio $\sigma_0/\sigma$ for $\lambda=2$, a (common) coefficient of restitution $\al=\al_0=0.8$, and a (reduced) density $n\sigma^2=0.1$. It is quite apparent that the agreement between theory and both MD and DSMC simulations is excellent, with deviations appearing when the size ratio is large. As before, although not shown here, the agreement deteriorates with large contrast ($\lambda\geq 4$) and high density where, as mentioned, the approximations made by the theory become to be less valid.

\begin{figure}
    \centering
    \includegraphics[width=1\linewidth]{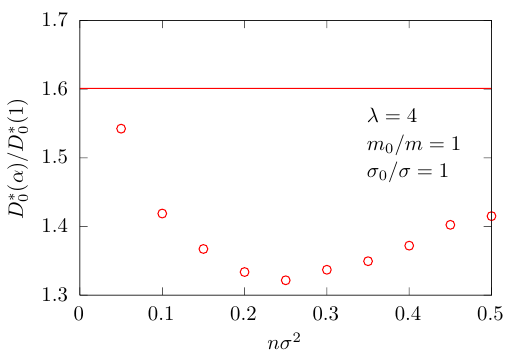}
    \caption{Comparison with MD simulations: case III. Normalized diffusion coefficient $D_0^*(\alpha)/D_0^*(1)$ as a function of the (reduced) density $n\sigma^2$ of the granular gas, for a tracer particle of equal mass and diameter of the particles of the granular gas. The (common) coefficient of restitution is $\alpha=\al_0=0.4$ and $\lambda\equiv\Delta_0/\Delta=4$. Theory in a solid line and MD results with open circles.}
    \label{fig.simcaseIII}
\end{figure}

Finally, in case III we study in more detail the density effects, which is done only using MD simulations as the DSMC method is not appropriate for this purpose. Specifically, Fig.~\ref{fig.simcaseIII} presents the normalized diffusion coefficient $D_0^*(\alpha)/D_0^*(1)$ as a function of the bath density $n\sigma^2$ for fixed coefficient of restitution and energy injection, and considering that particles are mechanically equivalent. In this situation, the theory predicts that there should be no density dependency on the scaled coefficient $D_0^*(\alpha)/D_0^*(1)$, while the MD results show that when increasing density there is an effect that can be as large as 20\%. Note that the case under consideration in the figure is rather extreme, with low restitution coefficients (very high inelasticity) and large value of the contrast in $\Delta$. Moderate cases do not present such strong deviation with density (not shown). These results are consistent with the hypothesis that the anomalous behavior of the MSD is due to correlations not included in the Enskog kinetic theory. These correlations play a relevant role on tracer diffusion at high densities and when the contrast between the tracer and the granular gas particles is large.

\section{Thermal diffusion segregation}
\label{sec7}

One of the most usual phenomenon appearing in multicomponent systems is the thermal diffusion segregation. It occurs in a non-convective steady state ($\mathbf{U}=\mathbf{0}$) due to the existence of a temperature gradient, which causes the movement of the different species of the mixture. In this situation, there is a balance between remixing of species caused by diffusion and segregation caused by temperature differences. The degree of segregation along the temperature gradient can be quantified by means of the so-called thermal diffusion factor $\Lambda$. In a steady state without convection ($\mathbf{U}=0$) and where the mass flux is zero ($\mathbf{j}_0=\mathbf{0}$), the thermal diffusion factor is defined as 
\beq
\label{6.1}
-\Lambda \frac{\partial \ln T}{\partial y}=\frac{\partial}{\partial y}\ln \Big(\frac{n_0}{n}\Big).
\eeq
Equation \eqref{6.1} has been simplified by assuming for the sake of simplicity that in the case of a two-dimensional system gradients occur only along the axis $y$. In addition, we also assume that the gravitational field is parallel to the thermal gradient, namely, $\mathbf{g}=-g\widehat{e}_y$, where $\widehat{e}_y$ is the unit vector in the positive direction of the $y$ axis.

\begin{figure}
\begin{center}
\includegraphics[width=1\linewidth]{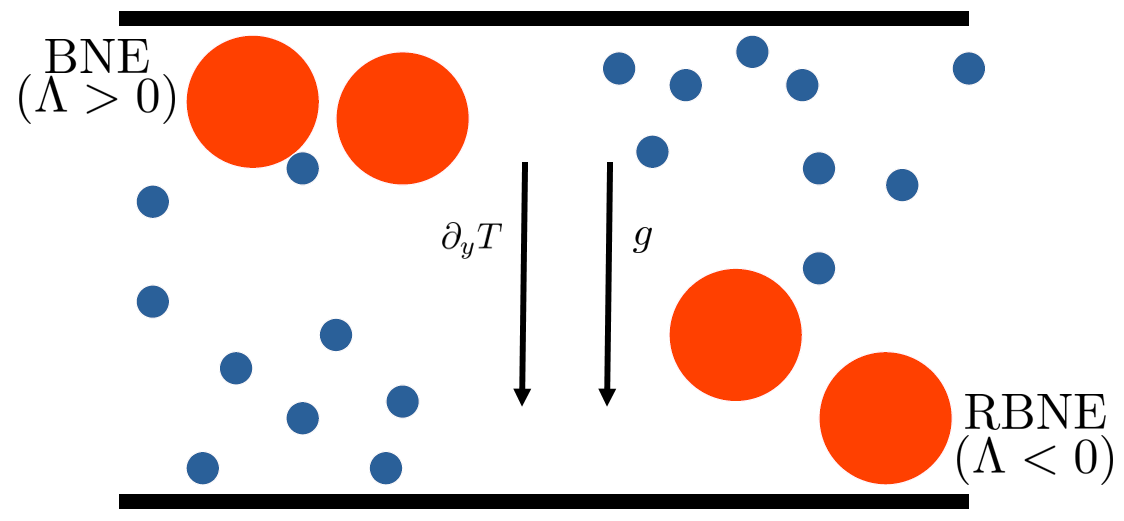}
\end{center}
\caption{Illustration of the segregation process behavior: small circles represent granular particles, and large circles represent impurities. The BNE ($\Lambda>0$) effect occurs when the impurity rises to the top, while the RBNE ($\Lambda<0$) effect occurs when the impurity sinks to the bottom of the system.\label{BNE}}
\end{figure}

Let us consider a scenario in which the tracer particles have a larger size than the granular gas particles ($\sigma_0>\sigma$). Furthermore, as said before, since gravity and the thermal gradient align, then the lower plate is hotter than the upper plate ($\partial_y\ln T<0$). Based on Eq.\ \eqref{6.1}, when $\Lambda>0$, \vicente{impurities (or intruders)} rise relative to granular gas particles ($\partial_z \ln (n_0/n)>0$), leading to an accumulation of tracer particles near the cooler plate, known as the Brazil nut effect (BNE). Conversely, for $\Lambda<0$, impurities descend compared to granular gas particles ($\partial_z \ln (n_0/n)<0$), resulting in an accumulation near the hotter plate, giving rise to the Reverse Brazil Nut Effect (RBNE). An illustrative example of the segregation process for a two-dimensional system is shown in Fig.\ \ref{BNE}. \RG{It is important to note that, although our study aims to capture the phenomenology of confined systems, we are modeling an unconfined two-dimensional system where collisions are described by the $\Delta$-model.}

Let us determine the thermal diffusion factor. In the steady state, assuming that $j_{0,y}^{(1)}=0$ and $\mathbf{U}=\mathbf{0}$, the momentum balance equation \eqref{1.12} reduces to $\partial_y p=-\rho g$. In dimensionless form, this relationship can be expressed as
\beq
\label{6.2}
p^*+\left(p^*+\phi \frac{\partial p^*}{\partial \phi}\right)\frac{\partial_y \ln n}{\partial_y \ln T}=-\Bigg({g}^*\vicente{-\frac{1}{2}\Delta^*\frac{\partial p^*}{\partial \Delta^*}}\Bigg),
\eeq
where ${g}^*=\rho {g}/n \partial_y T<0$ is a dimensionless parameter measuring the gravity relative to the thermal gradient \vicente{and $\frac{1}{2}\Delta^*\partial_{\Delta^*}p^*=\sqrt{\pi/2}\chi \phi \Delta^*$ for $d=2$. According to Eq.\ \eqref{0.1}, the condition $j_{0,y}^{(1)}=0$ leads to the relation 
\beq
\label{6.2.1}
-D_0^*\partial_y \ln n_0=D^*\partial_y \ln n+D_T^*\partial_y \ln T.
\eeq
From Eqs.\ \eqref{6.2} and \eqref{6.2.1}, the thermal diffusion factor $\Lambda$ can be expressed in terms of the dimensionless diffusion coefficients $D_0^*$, $D^*$, and $D_T^*$ as
\beqa
\label{6.3}
\Lambda&=&\frac{\partial_y \ln n}{\partial_y \ln T}-\frac{\partial_y \ln n_0}{\partial_y \ln T}\nonumber\\
&=&\frac{\xi D_T^{*}-\left(D_{0}^*+D^*\right)\left({g}^*+p^*-\frac{1}{2}\Delta^*\frac{\partial p^*}{\partial \Delta^*}\right)}{\xi D_{0}^*},
\eeqa
\nonumber\\
}
where $\xi=p^*+\phi \partial_\phi p^*$. It is quite apparent from Eq.\ \eqref{6.3} that the effect on the different parameters of the system (impurity plus granular gas) on the signature of $\Lambda$ is not simple at all. On the other hand, since Eq.\ \eqref{4.22} clearly shows that $D_0^*>0$, then the curves delineating the regimes between the
segregation toward the cold and the hot wall (BNE/RBNE transition) are determined from the condition
\beq
\label{6.4}
\xi D_T^{*}=\left(D_{0}^*+D^*\right)
\left({g}^*+p^*\vicente{-\frac{1}{2}\Delta^*\frac{\partial p^*}{\partial \Delta^*}}\right).
\eeq
To disentangle the different competing mechanisms appearing in the constraint \eqref{6.4} it is convenient to consider first some limiting situations where a more simple criterion can be derived.

\subsection{Mechanically equivalent particles}

In this limiting case, $D_T^*=0$, $D^*=-D_0^*$ and hence, Eq.\ \eqref{6.4} applies for any value of the
coefficients of restitution, masses, diameters, solid volume fraction, and $\Delta$ parameters. In
this case, as expected, no segregation appears in the system. 

\subsection{Low-density regime}

Let us assume now a \emph{dilute} granular gas ($n\sigma^2\to 0$) in the absence of gravity ($|{g}^*|=0$). In this limiting case, $p^*=\xi=1$ and the (dimensionless) diffusion transport coefficients can be easily identified from Eqs.\ \eqref{4.16}--\eqref{4.18}. According to these expressions, Eq.\ \eqref{6.4} yields the criterion
\beq
\label{6.5}
\widetilde{\Delta}^*\frac{\partial \gamma_0}{\partial \widetilde{\Delta}^*}\nu_D^*+\left(\gamma_0-\mu\right){\Delta}^*\frac{\partial \zeta^*}{\partial {\Delta}^*}=0.
\eeq
Equation \eqref{6.5} is still a quite complex relation in comparison with the one derived in the absence of gravity in the IHS model. \cite{GV09} In this latter case, the segregation criterion is simply given by $\gamma_0=\mu$.  
 
When the inhomogeneities in both the density and temperature can be neglected ($\partial_zT\to 0$) but gravity is nonzero, then $|{g}^*|\to \infty$ (thermalized systems). This situation (gravity dominates over thermal gradient) where segregation is essentially driven by gravity has been widely studied in simulations and experiments. \cite{HQL01,WHP01,BEKR03,SBKR05} In this limiting case, the segregation criterion in the low-density limit reads  
\beq
\label{6.6}
\gamma_0=\mu.
\eeq
Equation \eqref{6.6} agrees with the results obtained in the IHS when the granular gas is heated by means of an stochastic thermostat. \cite{G06}

\subsection{Moderately dense regime}

\subsubsection{Absence of gravity ($|{g}^*|\to 0$)}

Let us first consider a situation where gravity effects can be neglected ($|{g}^*|\to 0$). In this case, the condition $\Lambda=0$ yields the relation
\beq
\label{6.4a}
\xi D^*_T=\left(D_{0}^*+D^*\right)\left(p^*\RG{-\frac{1}{2}\Delta^*\frac{\partial p^*}{\partial \Delta^*}}\right).
\eeq
\RG{Let us look at what this equation means in simpler terms. When there is no gravity and the system is very dilute (meaning there are very few particles), the main factors that affect whether particles will separate (segregate) are the temperature gradient and the energy changes due to collisions and external energy inputs. Specifically, whether particles will rise or sink (BNE/RBNE transition) depends on the} balance of kinetic energy between the intruder and the granular gas. This balance is influenced by the energy loss from inelastic collisions (characterized by $\alpha$ and $\alpha_0$) and by the external energy inputs (characterized by $\Delta$ and $\Delta_0$).

Therefore, if the energy loss due to inelasticity and the energy input balance out for both species, no segregation will occur. However, if we modify the mechanical properties of each species, the number of collisions will vary. \RG{For example, if the impurity (or intruder) is larger than the other particles (i.e., $\omega\equiv \sigma_0/\sigma > 1$), it will experience more collisions because its collision frequency $\nu_0$ is proportional to its diameter $\sigma_0$. This leads to greater energy loss due to inelastic collisions, but it will also receive more energy from the external energy input $\Delta_0$}  [see Eq.\ \eqref{1.5} when $\Delta=\Delta_0$].

To find a region where segregation occurs, we need to break the symmetry between impurities and the particles of the granular gas. For (dry) granular gases, this can be achieved by modifying the mechanical properties of the mixture. For example, larger particles experience more collisions, leading to greater energy loss. When the impurity is larger, it undergoes more collisions, `cools down', and moves to the bottom wall (RBNE) since no extra velocity is added.\cite{G08a,GV12,G19} In granular suspensions, the thermostat discriminates the interaction with species based on their mechanical properties, thus promoting segregation. \cite{GG23}

\begin{figure}
\begin{center}
\begin{tabular}{lr}
\resizebox{6.5cm}{!}{\includegraphics{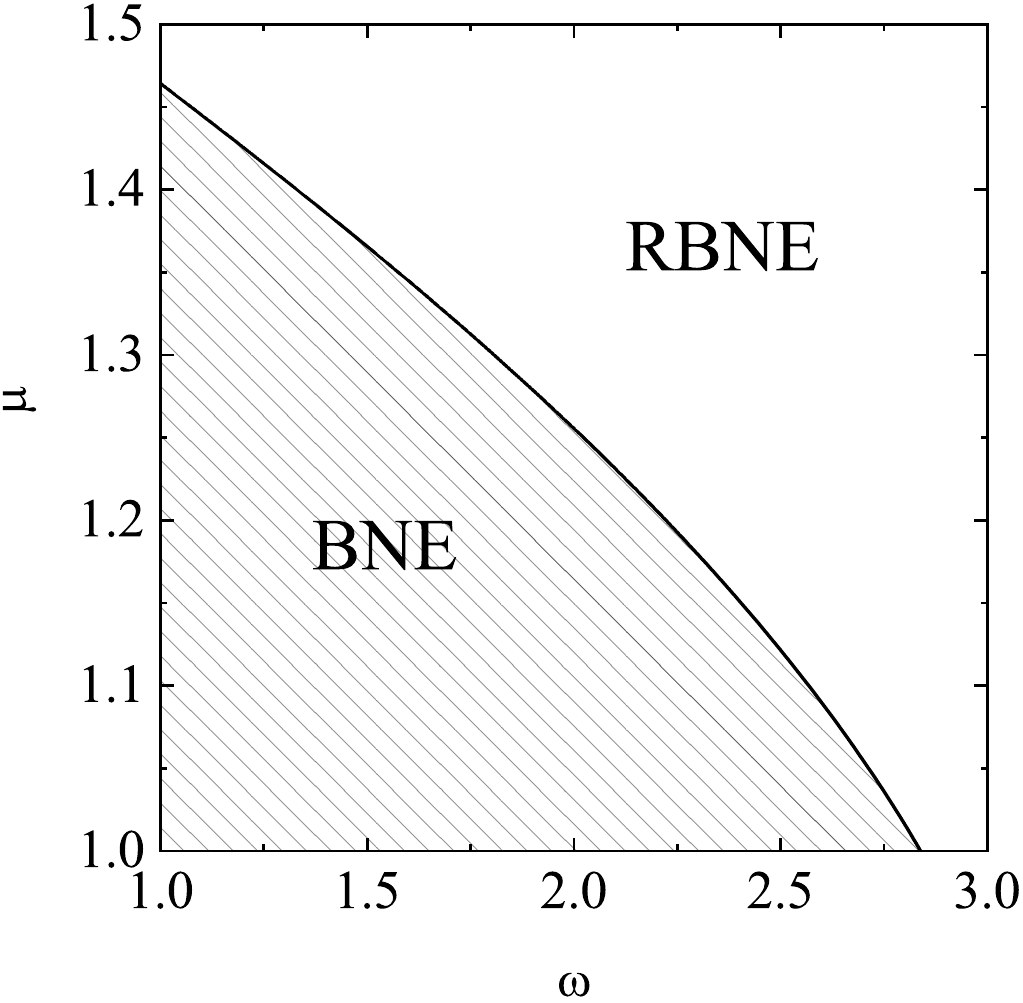}}
\end{tabular}
\end{center}
\caption{Plot of the marginal segregation curve ($\Lambda=0$) for a two-dimensional dilute ($n\sigma^2\to 0$) system  with a (common) coefficient of restitution $\alpha=\alpha_0=0.9$, and $|{g}^*|\to 0$. The points below the curve correspond to $\Lambda>0$ (BNE), while the points above the curve correspond to $\Lambda<0$ (RBNE). Here, $\Delta_0=\Delta/2$.\label{fig_seg1}}
\end{figure}

In this work, we will break the symmetry by ensuring that the larger species (the impurity or intruder), which experiences more collisions, receives less energy input per collision. We will achieve this by setting a common coefficient of restitution ($\alpha = \alpha_0$) and imposing $\Delta_0 < \Delta$. Therefore, for a fixed mass ratio $\mu\equiv m_0/m$ and a sufficiently large diameter ratio $\omega$, the intruder will lose energy and migrate to the bottom wall (RBNE). This effect can be observed in Fig.\ \ref{fig_seg1}, where we illustrate a dilute two-dimensional system ($n\sigma^2=0$) with a common coefficient of restitution coefficient $\alpha=0.9$, and $\Delta_0=\Delta/2$. \RG{On the other hand, if we keep the size ratio $\omega$ the same and change the mass ratio $\mu$, we see that as the intruder gets heavier, it moves from rising to sinking (transition from BNE to RBNE).}
This result contrasts with those found in Ref.\ \onlinecite{GMV13a} for dilute granular mixtures, although it is important to note that the cited study considered arbitrary molar fractions.

As we decrease the contrast of energy parameter $\lambda$ ($\Delta_0=\lambda\Delta$), we intensify the symmetry breaking, apparently reinforcing the segregation criterion. However, the intruder loses so much energy that it is unable to penetrate the sea of granular particles near the upper (cold) plate, which move much faster, and eventually settles predominantly on the lower (hot) plate, resulting exclusively in RBNE (no segregation criterion can be found).

\begin{figure}
\begin{center}
\begin{tabular}{lr}
\resizebox{6.5cm}{!}{\includegraphics{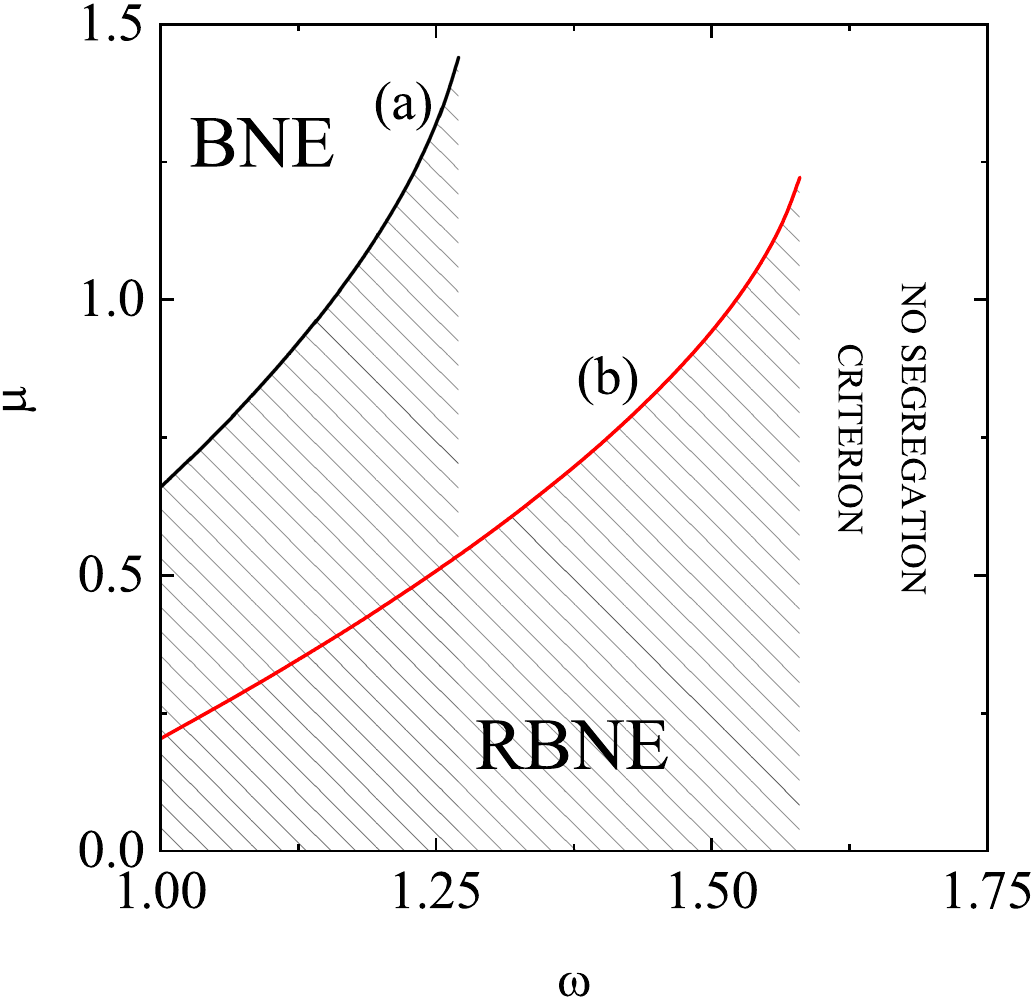}}
\end{tabular}
\end{center}
\caption{Plot of the marginal segregation curve ($\Lambda=0$) for $d = 2$, $\alpha_0=0.9$, $|{g}^*|\to 0$,  $\phi=0.1$, and two values of $\alpha$: 0.9 (a) and 0.6 (b). The points below the curve correspond to $\Lambda<0$ (RBNE), while the points above the curve correspond to $\Lambda>0$ (BNE). Here, $\Delta_0=\Delta/2$.
\label{fig_seg1b}}
\end{figure}

Let us now analyze the effect of density [or equivalently, the solid volume fraction $\phi=(\pi/4)n\sigma^2$] on segregation. As density increases, the frequency of collisions \RG{rises}. For instance, if we maintain the same parameters as depicted in Fig. \ref{fig_seg1} but increase the density (e.g., by setting $\phi=0.1$), locating a region where $\Lambda$ cancels becomes more challenging. This phenomenon mirrors the difficulty observed in the dilute case when $\lambda$ is too small. \RG{To expand the region where the BNE/RBNE transition occurs as we increase $\omega$, we need to reduce the impact of collisions and energy loss in the intruder. We do this by reducing $\alpha$ while keeping $\alpha_0=0.9$.} This expansion is evident in Fig.\ \ref{fig_seg1b}, where we have plotted two values of $\alpha$: 0.6 and 0.9. However, the effect of density now causes that the disparity in particle density between cold and hot walls become increasingly apparent. \RG{This means the intruder gets more collisions in the cold region, which can counteract the effect of extra collisions from the higher energy particles in the hot zone. As a result, this can flip the RBNE and BNE regions, as shown in Fig.\ \ref{fig_seg1b}.} This marginal segregation curve qualitatively agrees with those obtained for moderately dense granular and molecular gases, with and without interstitial fluid,\cite{G08a,GG23} allowing for similar conclusions about the effect of density on segregation.

\subsubsection{Thermalized systems ($\partial_z T\to 0$)}

\begin{figure}
\begin{center}
\begin{tabular}{lr}
\resizebox{6.5cm}{!}{\includegraphics{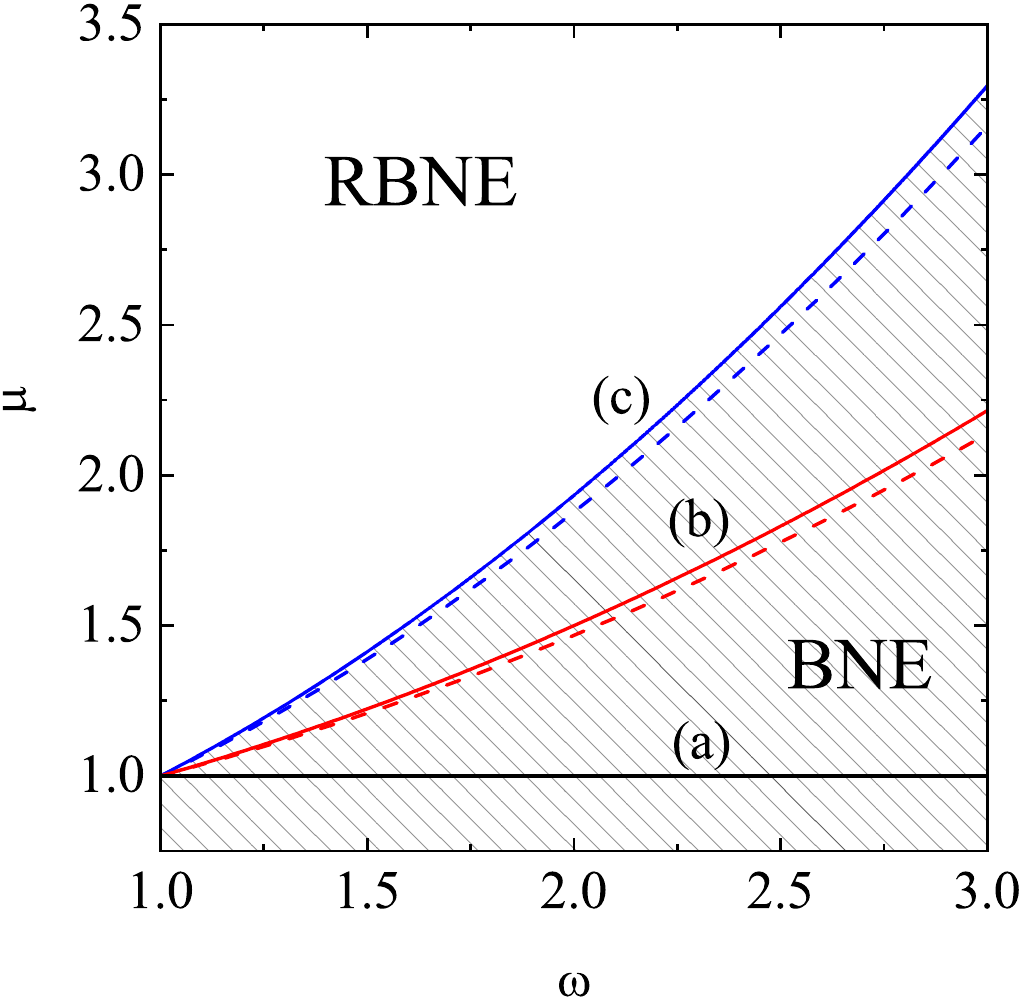}}
\end{tabular}
\end{center}
\caption{Plot of the marginal segregation curve ($\Lambda=0$) for $d = 2$, $\alpha=\alpha_0=0.8$ (solid lines) and 1 (dashed lines), $|{g}^*|\to \infty$, and three different values of the solid volume fraction $\phi$: $\phi=0$ (a), $\phi=0.1$ (b), and $\phi=0.2$ (c). The points below the curve correspond to $\Lambda>0$ (BNE), while the points above the curve correspond to $\Lambda<0$ (RBNE). Here, $\Delta=\Delta_0$.\label{fig_seg2}}
\end{figure}

We will now examine a scenario where the segregation dynamics are primarily influenced by the gravitational force. In such cases, $|{g}^*|\to \infty$, making the temperature gradient negligible ($\partial_y T\to 0$) and the condition $\Lambda=0$ yields the relation
\beq
\label{6.5b}
D_{0}^*+D^*=0.
\eeq
Obtaining thermalized systems is achievable in experimental setups and numerical simulations that involve shaken or sheared systems. \cite{HQL01,BEKR03,SBKR05,QDDWJZZ21} Figure \ref{fig_seg2} illustrates the segregation criterion for both a granular gas ($\alpha=\alpha_0=0.8$) and a molecular one ($\alpha=\alpha_0=1$). In this latter case, $\Delta=\Delta_0=0$ so that, we recover the segregation results for undriven granular mixtures. \cite{G11} Three different values of the volume fraction $\phi$ are considered in Fig.\ \ref{fig_seg2}: $\phi=0$, $\phi=0.1$, and $\phi=0.2$. Similar to driven granular mixtures (see Figs. 4 and 6 of Ref.\ \onlinecite{G09} and Figs. 13 and 14 of Ref.\ \onlinecite{GG23}), we observe a complete reversal of the RBNE/BNE transition when gravity is introduced, contrasting the phase diagrams shown in Fig.\ \ref{fig_seg1b} for finite densities. Furthermore, both the effects of inelasticity and the injection of energy introduced by $\Delta$ and $\Delta_0$ are suppressed by gravity, mimicking the behavior observed in granular suspensions and driven granular gases. \cite{G09,GG23} 

To provide an explanation for the BNE/RBNE transition, we can consider the interplay between particle mass and collision frequency. When intruder particles are heavier, their mass makes it easier for them to settle to the bottom of the container due to the effect of gravity, favoring the RBNE effect. Conversely, when the size ratio $\omega$ increases, tracer particles experience a higher number of collisions. These collisions induce a ``buoyancy-like" effect on the intruder, driven by the pressure exerted by surrounding granular particles. Consequently, the intruder is effectively lifted against gravity, leading to the observation of the BNE effect. On the other hand, the effect of density, characterized by the volume fraction $\phi$, is to increase the buoyancy effect, thus expanding the BNE region.

\subsubsection{General case}

\begin{figure}
\begin{center}
\begin{tabular}{lr}
\resizebox{6.5cm}{!}{\includegraphics{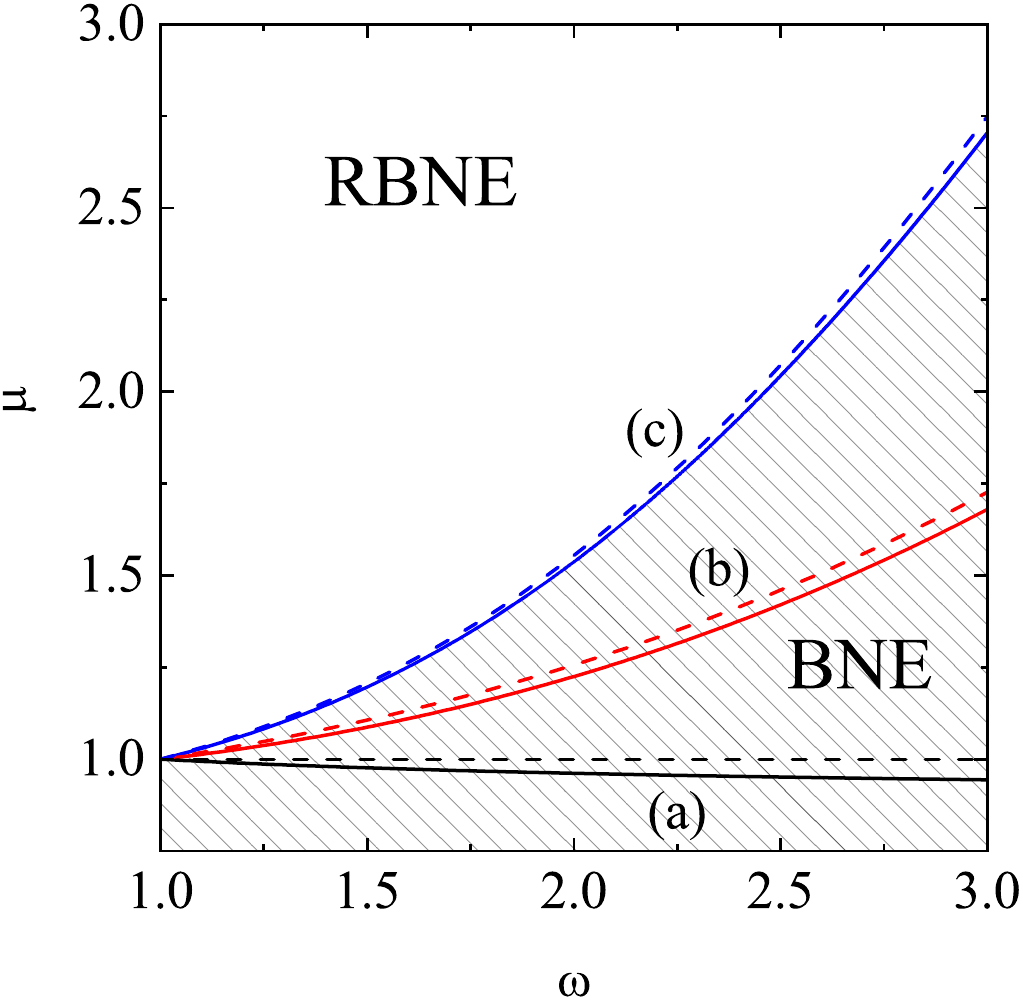}}
\end{tabular}
\end{center}
\caption{ Plot of the marginal segregation curve ($\Lambda=0$) for $d = 2$, $\alpha=\alpha_0=0.8$ (solid lines) and 1 (dashed lines), $|{g}^*|=1$, and three different values of the solid volume fraction $\phi$: $\phi=0$ (a), $\phi=0.1$ (b), and $\phi=0.2$ (c). The points below the curve correspond to $\Lambda>0$ (BNE), while the points above the curve correspond to $\Lambda<0$ (RBNE). Here, $\Delta=\Delta_0$.\label{fig_seg3}}
\end{figure}

Finally, we examine the general case where the impact of the temperature gradient is comparable to that of gravity. To exemplify this, Fig. \ref{fig_seg3} displays the marginal segregation curve for a reduced gravity $|{g}^*|=1$ and for the same systems as plotted in Fig.\ \ref{fig_seg2}.

The primary conclusion to be drawn is that, similar to dry granular mixtures and granular suspensions,\cite{G19,GG23} the effect of gravity is significantly more pronounced than that of the thermal gradient, as the BNE/RBNE transition is maintained. Additionally, the influence of inelasticity is also notably attenuated. Furthermore, in alignment with thermalized cases, an increase in density enlarges the region where the BNE effect is observed. Since ordinary temperature plays no role in the motion of grains, at the core of statistical description we can establish an equivalence between the density and the external driving in terms of the kinetic energy provided to the particles. Therefore, an analogy can be drawn between adjusting density and altering the shaking strength in experimental setups.\cite{TAH03,ATH06} Specifically, decreasing the shaking strength expands the BNE region. This phenomenon has already been observed in freely evolving and driven granular mixtures, as well as in granular suspensions. \cite{G19,GG23}

\section{Discussion}
\label{sec8}

In this work, we have considered a kinetic theory approach to determine the diffusion transport coefficients for a granular
binary mixture in the tracer limit. The injection of energy in the system is accounted for by a collisional model, the so-called $\Delta$-model. \cite{BRS13} In particular, in each collision an amount of relative velocity $\Delta$ is added to the velocities of the colliding particles. This energy injection acts as an effective thermostat, allowing the system to reach a steady state. \cite{BGMB13,BMGB14} 

It should be noted that in a granular binary mixture (composed of two types of particles differing in mass, diameter, inelasticity, or the value of $\Delta$ at collisions), the steady state of the $\Delta$-model is characterized by a temperature ratio $T_1/T_2$ different from 1 (breakdown of the energy equipartition). Although the granular temperature $T$ is the relevant one at a hydrodynamic level, the effect of the energy nonequipartititon on transport must be accounted for since its impact on transport is generally significant, as has been clearly shown in previous works for IHS [see for instance, the review \onlinecite{ChGG22}]

In the tracer limit, since the state of the granular gas (excess component) is not affected by the presence of tracer or impurity particles, the momentum and heat fluxes of the system (impurities plus granular gas) are the same as that for the granular gas. In this limiting case the mass flux of the impurities $\mathbf{j}_0$ is the relevant flux of the problem, and therefore the goal of this paper has been to determine $\mathbf{j}_0$ in terms of the parameter space of the system. 

The theoretical framework of our study is based on the Enskog kinetic equation, which extends the Boltzmann results for low-density gases to higher densities. Under the tracer condition, the kinetic equation for the velocity distribution function $f$ of the granular gas decouples from the tracer velocity distribution function $f_0$, and thus the distribution $f$ obeys the usual nonlinear Enskog equation. In addition, since the collisions between the tracer particles themselves can be neglected, the distribution $f_0$  obeys a linear Enskog--Lorentz kinetic equation.

In the Navier–Stokes domain, the constitutive equation for the mass flux $\mathbf{j}_0^{(1)}$ is given by Eq.\ \eqref{0.1}, where the superscript $1$ in $\mathbf{j}_0$ means that the flux is computed to first order in the spatial gradients of $n_0$, $n$, and $T$. Equation \eqref{0.1} includes three transport coefficients: (a) the tracer diffusion coefficient, $D_0$, associated with the gradient of tracer concentration $\nabla n_0$, (b) the mutual diffusion coefficient, $D$, related with the diffusion of impurities into the bulk particles; it is associated with the density gradient $\nabla n$, and (c) the thermal diffusion coefficient, $D_T$, associated with the temperature gradient $\nabla T$. These three coefficients have been obtained here in the so-called first Sonine approximation, with results expressed in Eqs.~\eqref{4.4}-\eqref{4.6}. While the first Sonine approximation simplifies the solution of the coupled integral equations, higher order terms in the Sonine expansion may be necessary to achieve greater accuracy, especially in extreme conditions such as  very strong inelasticity and/or very disparate mass and diameter ratios. 

To assess the validity of the analytical predictions, numerical simulations were performed for the tracer diffusion coefficient, $D_0$. Its value is obtained by differentiating, with respect to time, the mean square displacement of the tracer particles at long times, when the diffusive limit is reached.
Two simulation techniques have been employed: the DSMC method \cite{B94,MS96} and MD simulations using the event-driven method. \cite{TS05,AT17,FS23}
The DSMC method provides ``exact'' solutions of the kinetic equation beyond the first Sonine polynomial approximation. Thus, deviations of DSMC simulations from the (approximate) analytical results could in principle be mitigated by considering higher-order Sonine corrections to the distribution functions $f$ and $f_0$. MD simulations, however, do not rely on the assumptions of the Enskog kinetic equation, such as the molecular chaos assumption, the form of the pair correlation function or any other correlations ignored in the theoretical description.

It is interesting to note that the MD simulations show an intermediate regime between the short-time (ballistic) and the long-time (diffusive) regimes (see Fig.\ \ref{fig.MSD}); it deserves further study. This new regime is particularly noticeable when the dynamic properties of the impurity are very different from those of the bulk (granular gas particles); in particular, when the energy input to the tracer particles is greater than that to the particles of the granular gas.

The results show that the agreement between the analytical and numerical results is excellent for moderate densities and not very strong inelasticities, as expected. However, as the dissipation and density increase, some discrepancies between simulations and theoretical results appear. The discrepancy between theory and DSMC simulations is rather small and has its origin in the first Sonine approximation considered in the theoretical development. MD simulations, however, show larger deviations from the theoretical results, which may be related to the breakdown of the molecular chaos hypothesis, especially at high densities.

As an application of the derived expressions for the 
diffusion transport coefficients, the thermal diffusion factor $\Lambda$ [defined in Eq.\ \eqref{6.1}] was obtained. The evaluation of $\Lambda$ provides new insights into the segregation phenomenon in dense granular mixtures in the presence of gravity and/or a temperature gradient. The phase diagrams obtained for the transition between the BNE and the RBNE\cite{RSPS87,DRC93} provide a more comprehensive understanding of how variations in system parameters such as mass, size and inelasticity control segregation. The results suggest that segregation dynamics in confined granular systems differ significantly from those in driven or freely cooling systems of IHS,\cite{BEGS08,G08a,GV09,G09} reiterating the fact that energy input can influence segregation criteria. On the other hand, a similar behavior has been observed in some situations with the phase diagrams of segregation obtained for granular suspensions. \cite{GG23}

\vicente{The results derived here have been obtained in the context of a very simple model where the coefficients of restitution $\al$ and $\al_0$ are constant. However, several experiments and simulations have shown that the coefficients of restitution depend on the relative velocity of the colliding spheres. \cite{BP04} Some works \cite{BP03,BSSP04} devoted to these viscoelastic models have shown that the Navier--Stokes transport coefficients exhibit qualitative differences from the ones obtained with the simplifying assumption of a constant coefficient of restitution. In particular, if the granular gas is kept at constant temperature, the approximation of an effective coefficient of restitution can be applicable. \cite{DBPB13}  However, in the case of the presence of a temperature gradient, the coefficients of restitution will vary in different parts of the system and hence the segregation criterion will be different as the one obtained in Sec.\ \ref{sec7}. 

\RG{Incorporating a velocity-dependent coefficient of restitution into theoretical studies is very challenging due to the increased complexity of the mathematical treatment required.} Thus, it would be interesting to extend the present results to the case of viscoelastic particles in order to see if the behaviors found here for the diffusion transport coefficients are also present (at least from a qualitative level) when the collisions are described by an impact velocity dependent coefficients of restitution.      
}

\RS{The $\Delta$-model qualitatively describes the effective two-dimensional dynamics of vibrofluidized three dimensional granular systems in shallow boxes, but if fails to describe the density instability that has been observed in experiments and simulations~\cite{}. The origin of this difference is the use of a constant value of $\Delta$, which results in a pressure that grows monotonically with density. Extensions of the model, where $\Delta$ increases with time, adequately reproduce the instability.~\cite{RSG18} The associated kinetic theory is more complex but similar approaches as those presented here could be used to investigate diffusion in this case.}

Despite the success of the current theoretical framework, future work could focus on refining the analytical expressions for the diffusion coefficients, perhaps by using alternative methods for solving the Enskog-Lorentz equation or by exploring higher order terms in the Sonine polynomial expansion. In addition, the interplay between confinement, inelasticity, and particle shape could be further investigated, as these factors are likely to influence transport properties in granular gases under more complex conditions.

In summary, the present study provides a comprehensive theoretical treatment of mass transport in a dense granular gas within a confined geometry, providing valuable predictions for tracer diffusion coefficients. The results highlight the usefulness of the $\Delta$ model in mimicking energy transfer between the horizontal and vertical degrees of freedom, and the comparison with simulation data confirms the validity of the theoretical approach. The work also extends previous findings by introducing a segregation criterion that distinguishes between BNE and RBNE under a thermal gradient and gravity, contributing to a broader understanding of granular segregation in confined systems.

\acknowledgments 

The work of R.G.G. and V.G. is supported from Grant No. PID2020-112936GB-I00 funded by MCIN/AEI/ 10.13039/501100011033. The work of R.B. is supported from Grant Number PID2020-113455GB-I00 and PID2023-147067NB-I00.
The work of R.S. is supported by the Fondecyt Grant No.~1220536 and Millennium Science Initiative Program NCN19\_170 of ANID, Chile. 
\vspace{0.5cm}

\textbf{AUTHOR DECLARATIONS}\\
\textbf{Conflict of Interest}\\
The authors have no conflicts to disclose.

\vspace{0.5cm}
\textbf{Author Contributions} \\
\textbf{Rub\'en G\'omez Gonz\'alez}: Formal analysis (equal); Investigation (equal); Writing–review\&editing (equal). \textbf{Vicente Garzo}: Formal analysis (equal); Investigation (equal); Writing/Original Draft Preparation (lead); Writing–review\&editing(equal). \textbf{Ricardo Brito}: Conceptualization (equal); Investigation (equal); Software (lead); Writing–review\&editing(equal). \textbf{Rodrigo Soto}: Conceptualization(equal); Investigation (equal); Software (lead); Writing–review\&editing (equal).

\vspace{0.5cm}

\textbf{DATA AVAILABILITY}\\

The data that support the findings of this study are available from the corresponding author upon reasonable request.

\appendix
\section{First Sonine approximation to the diffusion transport coefficients}
\label{appA}     

\begin{widetext}
In this Appendix we provide some technical details on the evaluation of the diffusion transport coefficients by considering the first Sonine approximations \eqref{4.1}--\eqref{4.3}. Let us evaluate each transport coefficient separately. 

\subsection{Tracer diffusion coefficient $D_0$}

We consider first the tracer diffusion coefficient $D_0$. In this case, $\boldsymbol{\mathcal{B}}_0$ is given by Eq.\ \eqref{4.2} and Eq.\ \eqref{3.7} yields
\beq
\label{a1}
d\frac{m_0 \rho_0}{\rho}\zeta^{(0)}T\partial_T D_0+\frac{m_0^2}{\rho T_0} D_0\int d\mathbf{v} m_0 \mathbf{V}\cdot J_0^{(0)}\Big[f_{0\text{M}} \mathbf{V},f^{(0)}\Big]=-d n_0 T_0^{(0)},
\eeq
where $\rho_0=m_0n_0$ and use has been made of the result
\beq
\label{a2}
\int d\mathbf{v}\; m_0 \mathbf{V}\cdot \mathbf{B}_0=-d n_0 T_0^{(0)}.
\eeq
Equation \eqref{a1} can be rewritten in terms of the dimensionless coefficient $D_0^*=(m_0^2\nu/\rho T)D_0$ as
\beq
\label{a3}
-\frac{1}{2}\zeta^* D_0^* \Big(1-\widetilde{\Delta}^*\frac{\partial \ln D_0^*}{\partial \widetilde{\Delta}^*}\Big)+\nu_\text{D}^*D_0^*=\gamma_0,
\eeq
where we recall that $\zeta^*=\zeta^{(0)}/\nu$, $\gamma_0=T_0^{(0)}/T$ and 
\beq
\label{a4}
\nu_\text{D}^*=-\frac{1}{dn_0T_0^{(0)}\nu}\int d\mathbf{v} m_0 \mathbf{V}\cdot J_0^{(0)}\Big[f_{0\text{M}} \mathbf{V},f^{(0)}\Big].
\eeq
Upon obtaining Eq.\ \eqref{a3}, we have taken into account that
\beq
\label{a5}
T\partial_T D_0^*=-\frac{1}{2}\widetilde{\Delta}^*\frac{\partial D_0^*}{\partial \widetilde{\Delta}^*}.
\eeq
The expression of $\nu_\text{D}^*$ was estimated in Ref.\ \onlinecite{GBS21} when the zeroth-order distribution $f^{(0)}(\mathbf{V})$ of the granular gas is approximated by the Maxwellian distribution \eqref{1.15} with the replacement $\mathbf{v}\to \mathbf{V}$. In this approximation, one gets the result
\beq
\label{a6}
\nu_\text{D}^*=\frac{2\pi^{\frac{d-1}{2}}}{d\Gamma\left(\frac{d}{2}\right)}\chi_0 \left(\frac{\bar{\sigma}}{\sigma}\right)^{d-1} M \Bigg[\Big(\frac{1+\theta}{\theta}\Big)^{1/2}(1+\al_0)+\sqrt{\pi}\Delta_0^*\Bigg].
\eeq
In the HSS, $\zeta^*=0$ and Eq.\ \eqref{a3} gives the expression \eqref{4.4} for $D_0^*$.

\subsection{Thermal diffusion coefficient $D_T$}

In a similar way as $D_0$, the thermal diffusion coefficient $D_T$ can be easily obtained from Eqs.\ \eqref{3.6} and \eqref{4.1} as
\beq
\label{a7}
d\rho \zeta^{(0)}T\partial_T D_T+\frac{1}{2}d\rho \zeta^{(0)}\Big(1-\Delta^*\frac{\partial \ln \zeta^*}{\partial \Delta^*}\Big)D_T-d\rho \nu \nu_\text{D}^* D_T=\int d\mathbf{v}\; m_0 \mathbf{V}\cdot \mathbf{A}_0.
\eeq
According to the expression \eqref{3.9} of $\mathbf{A}_0$, the right hand side of Eq.\ \eqref{a7} can be computed as
\beq
\label{a8}
\int d\mathbf{v}\; m_0 \mathbf{V}\cdot \mathbf{A}_0=-d n_0 T\Big(\gamma_0-\frac{1}{2}\widetilde{\Delta}^*\frac{\partial \gamma_0}{\partial \widetilde{\Delta}^*}\Big)+d \frac{\rho_0}{\rho}p\Big(1-\frac{1}{2}\Delta^*\frac{\partial \ln p^*}{\partial \Delta^*}\Big)-\int d\mathbf{v}\; m_0 \mathbf{V}\cdot \boldsymbol{\mathcal{K}}_0\Big[T \frac{\partial f^{(0)}}{\partial T}\Big],
\eeq
where
\beq
\label{a9}
\Delta^*\frac{\partial p^*}{\partial \Delta^*}=\frac{2^d}{\sqrt{2\pi}}\chi \phi \Delta^*.
\eeq
Equation \eqref{a7} can be rewritten in terms of the dimensionless coefficient $D_T^*=(\rho \nu/n_0 T)D_T$ as
\beqa
\label{a10}
& & \frac{1}{2}\zeta^*D_T^*\Big(1-\widetilde{\Delta}^*\frac{\partial \ln D_T^*}{\partial \widetilde{\Delta}^*}\Big)+\frac{1}{2}\zeta^* D_T^*\Big(1-\Delta^*\frac{\partial \ln \zeta^*}{\partial \Delta^*}\Big)-\nu_\text{D}^*D_T^*=-\gamma_0\Big(1-\frac{1}{2}
\widetilde{\Delta}^*\frac{\partial \ln \gamma_0}{\partial \widetilde{\Delta}^*}\Big)+\frac{m_0}{m}\Big(p^*-\frac{2^{d-1}}{\sqrt{2\pi}}\chi \phi \Delta^*\Big)
\nonumber\\
& & -
\frac{1}{d n_0 T}\int d\mathbf{v}\; m_0 \mathbf{V}\cdot \boldsymbol{\mathcal{K}}_0\Big[T \frac{\partial f^{(0)}}{\partial T}\Big].
\eeqa
The collision integral involving the operator $\mathcal{K}_0$ has been evaluated in the Appendix \ref{appB} with the result
\beq
\label{a11}
\frac{1}{d n_0 T}\int d\mathbf{v}\; m_0 \mathbf{V}\cdot \boldsymbol{\mathcal{K}}_0\Big[T \frac{\partial f^{(0)}}{\partial T}\Big]=2^d \left(\frac{\bar{\sigma}}{\sigma}\right)^{d} \phi \chi_0 M_0 \Bigg[\frac{1+\al_0}{2}+\frac{\Delta_0^*}{\sqrt{\pi}}
\left(\frac{\theta}{1+\theta}\right)^{1/2}\Bigg].
\eeq
In the HSS ($\zeta^*=0$), the differential equation \eqref{a10} becomes an algebraic equation whose solution gives Eq.\ \eqref{4.5} for $D_T^*$.

\subsection{Diffusion coefficient D}

In the case of the diffusion coefficient $D$, $\boldsymbol{\mathcal{C}}_0$ is given by Eq.\ \eqref{4.3}. As for $D^*$ and $D_T^*$, the equation for the dimensionless diffusion coefficient $D_0^*$ can be derived from Eq.\ \eqref{4.8} (with $\boldsymbol{\mathcal{C}}=\mathbf{0}$) by multiplying both sides of this equation by $m_0 \mathbf{V}$ and integrating on velocity. The result is
\beq
\label{a12}
\frac{1}{2}\zeta^*D^*\Big(1-\widetilde{\Delta}^*\frac{\partial \ln D^*}{\partial \widetilde{\Delta}^*}\Big)-\nu_D^* D^*=
\frac{1}{dn_0 T}\int d\mathbf{v}\; m_0 \mathbf{V}\cdot \mathbf{C}_0-\zeta^*\Big(1+\phi\frac{\partial \ln \chi}{\partial \phi}\Big)D_T^*,
\eeq
where $D^*=(m_0\nu/n_0 T)D$. The first term on the right-hand side of Eq.\ \eqref{a12} is given by
\beq
\label{a13}
\frac{1}{d n_0 T}\int d\mathbf{v}\; m_0 \mathbf{V}\cdot \mathbf{C}_0=-\phi \frac{\gamma_0}{\partial \phi}+\mu \left(p^*+\phi \frac{\partial p^*}{\partial \phi}\right)-\frac{(1+\omega)^{-d}}{\chi_0 T}\Big(\frac{\partial \mu_0}{\partial \phi}\Big)_{T,n_0}\frac{1}{dn_0T}\int d\mathbf{v}\; m_0 \mathbf{V}\cdot \boldsymbol{\mathcal{K}}_0\Big[f^{(0)}\Big].
\eeq
In Eq.\ \eqref{a13}, we recall that $\mu=m_0/m$ and $\omega=\sigma_0/\sigma$. The collisional integral appearing in the last term on the right hand side of Eq.\ \eqref{a13} has been also computed in the Appendix \ref{appB} with the result
\beq
\label{a14}
\frac{1}{d n_0 T}\int d\mathbf{v}\; m_0 \mathbf{V}\cdot \boldsymbol{\mathcal{K}}_0\Big[f^{(0)}\Big]=2^d \left(\frac{\bar{\sigma}}{\sigma}\right)^{d} \phi \chi_0 M_0 \left(\frac{1+\theta}{\theta}\right)\Bigg[\frac{1+\al_0}{2}+\frac{2\Delta_0^*}{\sqrt{\pi}}
\left(\frac{\theta}{1+\theta}\right)^{1/2}\Bigg].
\eeq
The expression \eqref{4.6} for $D^*$ in the HSS can be easily obtained when one takes $\zeta^*=0$ in Eq.\ \eqref{a12} and takes into account Eqs.\ \eqref{a13} and \eqref{a14}.

\subsection{Derivatives with respect to $\Delta^*$ and $\widetilde{\Delta}^*$}

It is readily apparent that the diffusion transport coefficients are given in terms of some derivatives. Let us evaluate these derivatives in the HSS. We start with the derivative $\partial \gamma_0/\partial \phi$ in the steady state ($\zeta^*=\zeta_0^*=0$). It can be obtained from the time-dependent equation for the temperature ratio $\gamma_0$:
\beq
\label{a15}
\zeta^* \widetilde{\Delta}^*\frac{\partial \gamma_0}{\partial \widetilde{\Delta}^*}=2\gamma_0 \left(\zeta^*-\zeta_0^*\right).
\eeq
In the steady state, Eq. \eqref{a15} is trivially satisfied. To determine the derivative $\partial \gamma_0/\partial \phi$ at the steady state, we take first the derivative with respect to $\phi$ in both sides of Eq.\ \eqref{a15} and then takes the steady state limit. Thus, one gets the identity
\beq
\label{a16b}
\frac{\partial \widetilde{\zeta}_0}{\partial \phi}=\frac{\partial \widetilde{\zeta}_0}{\partial \gamma_0}\frac{\partial \gamma_0}{\partial \phi}=0,
\eeq
where $\widetilde{\zeta}_0=\zeta_0^*/\chi_0$. Upon obtaining Eq.\ \eqref{a16b} use has been made of the fact that in the steady state $\partial \zeta^*/\partial  \phi=\zeta^*(\partial \ln \chi/\partial \phi)=0$ and 
$\widetilde{\zeta}_0=0$. Since in Eq.\ \eqref{a16b}, $\partial \widetilde{\zeta}_0/\partial \gamma_0$ is in general different from zero in the steady state, then one concludes that $\partial \gamma_0/\partial \phi=0$ in the steady state for the $\Delta$-model.

In addition, the derivative $\partial \gamma_0/\partial \widetilde{\Delta}$ can be also be obtained from Eq.\ \eqref{a15}. It is given by
\beq
\label{a16}
\widetilde{\Delta}^*\frac{\partial \gamma_0}{\partial \widetilde{\Delta}^*}=\gamma_0 \frac{\Delta^*\frac{\partial \zeta^*}{\partial {\Delta}^*}-\widetilde{\Delta}^*\left(\frac{\partial \zeta_0^*}{\partial \widetilde{\Delta}^*}\right)_{\gamma_0}}{\frac{1}{2}{\Delta}^*\frac{\partial \zeta^*}{\partial {\Delta}^*}+\gamma_0\left(\frac{\partial \zeta_0^*}{\partial \gamma_0}\right)_{\widetilde{\Delta}^*}},
\eeq
where
\beq
\label{a17}
\Delta^*\frac{\partial \zeta^*}{\partial {\Delta}^*}
=-\frac{\sqrt{2}\pi^{\frac{d-1}{2}}}{d\Gamma\left(\frac{d}{2}\right)} \chi(\phi) \Delta^*\left(4\Delta^{*}+\sqrt{2\pi}\al \right), \quad \widetilde{\Delta}^*\frac{\partial \zeta_0^*}{\partial \widetilde{\Delta}^*}=\Delta_0^*\frac{\partial \zeta_0^*}{\partial {\Delta_0^*}}.
\eeq

\section{Evaluation of the collision integrals}
\label{appB}

In this Appendix we compute the collision integrals appearing in the evaluation of the transport coefficients $D_T$ and $D$. Let us start with the collision integral
\beq
\label{b1}
I_{D_T}\equiv \int d\mathbf{v}\; m_0 \mathbf{V}\cdot \boldsymbol{\mathcal{K}}_0\Big[T \frac{\partial f^{(0)}}{\partial T}\Big]
\eeq
This integral can be computed by using the property
\beq
\label{b2}
\int d\mathbf{v}_1\; \psi(\mathbf{V}_1) \mathcal{K}_{0,i}[X(\mathbf{V}_2)]=-\bar{\sigma}^d \chi_0\int d\mathbf{V}_1 \int d\mathbf{V}_2\int d\widehat{\boldsymbol{\sigma}}
\Theta (\widehat{{\boldsymbol {\sigma }}}\cdot {\bf g}_{12})
(\widehat{\boldsymbol {\sigma }}\cdot {\bf g}_{12})\widehat{\sigma}_if_0^{(0)}(\mathbf{V}_1)X(\mathbf{V}_2)\Big[\psi(\mathbf{V}_1')-\psi(\mathbf{V}_1)\Big],
\eeq
where we recall that $\mathbf{V}_1'$ is defined by
\beq
\label{b3}
\mathbf{V}_1'=\mathbf{V}_1-M\left(1+\alpha_{0}\right)(\widehat{{\boldsymbol {\sigma }}}\cdot \mathbf{g}_{12})\widehat{{\boldsymbol {\sigma }}}-2M\Delta_{0}\widehat{{\boldsymbol {\sigma}}}.
\eeq
Using the property \eqref{b2}, the integral $I_{D_T}$ reads
\beqa
\label{b4}
I_{D_T}&=&-m_0\bar{\sigma}^d \chi_0\int d\mathbf{V}_1 \int d\mathbf{V}_2\int d\widehat{\boldsymbol{\sigma}}
\Theta (\widehat{{\boldsymbol {\sigma}}}\cdot {\bf g}_{12})
(\widehat{\boldsymbol {\sigma}}\cdot {\bf g}_{12})f_0^{(0)}(\mathbf{V}_1)\Big[T \frac{\partial f^{(0)}(\mathbf{V}_2)}{\partial T}\Big]
\widehat{{\boldsymbol {\sigma}}}\cdot \Big(\mathbf{V}_1'-\mathbf{V}_1\Big) \nonumber\\
&=&
m_0M\bar{\sigma}^d \chi_0\int d\mathbf{V}_1 \int d\mathbf{V}_2\int d\widehat{\boldsymbol{\sigma}}
\Theta (\widehat{{\boldsymbol {\sigma}}}\cdot {\bf g}_{12})
(\widehat{\boldsymbol {\sigma}}\cdot {\bf g}_{12})f_0^{(0)}(\mathbf{V}_1)\Big[T \frac{\partial f^{(0)}(\mathbf{V}_2)}{\partial T}\Big]
\Big[\left(1+\alpha_{0}\right)(\widehat{{\boldsymbol {\sigma }}}\cdot \mathbf{g}_{12})+2\Delta_{0}\Big].
\nonumber\\
\eeqa
In the hydrodynamic regime, the zeroth-order distribution function $f^{(0)}(\mathbf{V})$ of the granular gas has the scaled form
\beq
\label{b5}
f^{(0)}(\mathbf{V})=n v_0^{-d}\varphi(\mathbf{c},\Delta^*),
\eeq
where $\mathbf{c}=\mathbf{V}/v_0$ and the unknown scaled distribution $\varphi$ depends on $T$ through $\mathbf{c}$ and $\Delta^*$. According to Eq.\ \eqref{b5}, $f^{(0)}(\mathbf{V})$ has the property
\beq
\label{b6}
T \frac{\partial f^{(0)}}{\partial T}=-\frac{1}{2}\frac{\partial}{\partial \mathbf{V}}\cdot \left(\mathbf{V} f^{(0)}\right)-\frac{1}{2}n v_0^{-d}\Delta^*\frac{\partial \varphi}{\partial \Delta^*}.
\eeq
However, if one takes the Maxwellian approximation \eqref{1.15} for $f^{(0)}(\mathbf{V})$, then $\varphi \simeq \pi^{-d/2}e^{-c^2}$, and so
\beq
\label{b7}
T \frac{\partial f^{(0)}}{\partial T}\simeq-\frac{1}{2}\frac{\partial}{\partial \mathbf{V}}\cdot \left(\mathbf{V} f^{(0)}\right).
\eeq
For the sake of simplicity, the approximation \eqref{b7} is employed here to evaluate $I_{D_T}$. In this case,
\beqa
\label{b8}
I_{D_T}&=&
-\frac{1}{2}m_0M\bar{\sigma}^d \chi_0\int d\mathbf{V}_1 \int d\mathbf{V}_2f_0^{(0)}(\mathbf{V}_1)\frac{\partial}{\partial V_{2i}}\left(V_{2i} f^{(0)}(\mathbf{V}_2)\right)\int d\widehat{\boldsymbol{\sigma}}
\Theta (\widehat{{\boldsymbol {\sigma}}}\cdot {\bf g}_{12})
(\widehat{\boldsymbol {\sigma}}\cdot {\bf g}_{12})
\Big[\left(1+\alpha_{0}\right)(\widehat{{\boldsymbol {\sigma }}}\cdot \mathbf{g}_{12})+2\Delta_{0}\Big]\nonumber\\
&=&-\frac{1}{2}m_0M\bar{\sigma}^d \chi_0\int d\mathbf{V}_1 \int d\mathbf{V}_2f_0^{(0)}(\mathbf{V}_1)\frac{\partial}{\partial V_{2i}}\left(V_{2i} f^{(0)}(\mathbf{V}_2)\right)\Big[B_2\left(1+\alpha_{0}\right)g_{12}^2+2 B_1 \Delta_{0} g_{12}\Big],
\eeqa
where 
\beq
\label{b9}
B_k=\pi^{(d-1)/2}\frac{\Gamma\left(\frac{k+1}{2}\right)}{\Gamma\left(\frac{k+d}{2}\right)}.
\eeq
The integral $I_{D_T}$ can be split in two parts; one of them was already computed \cite{GHD07} when $\Delta_0=0$. Thus, the integral $I_{D_T}$ can be written as
\beq
\label{b10}
I_{D_T}=I_{D_T}^{(0)}+I_{D_T}^{(1)},
\eeq
where\cite{GHD07}
\beq
\label{b11}
I_{D_T}^{(0)}=d B_2 n \bar{\sigma}^d \chi_0 M_0 n_0 T (1+\al_0),
\eeq
and 
\beq
\label{b12}
I_{D_T}^{(1)}=
-B_1 m_0 M \bar{\sigma}^d \chi_0 \Delta_0 \int d\mathbf{V}_1 \int d\mathbf{V}_2f_0^{(0)}(\mathbf{V}_1)\frac{\partial}{\partial V_{2i}}\left(V_{2i} f^{(0)}(\mathbf{V}_2)\right)g_{12}.
\eeq
Integrating by  parts, Eq.\ \eqref{b11} can be written as
\beqa
\label{b13}
I_{D_T}^{(1)}&=&B_1 m_0 M \bar{\sigma}^d  \chi_0 \Delta_0 \int d\mathbf{V}_1 \int d\mathbf{V}_2f_0^{(0)}(\mathbf{V}_1)f^{(0)}(\mathbf{V}_2) V_{2i}  \frac{\partial g_{12}}{\partial V_{2i}} \nonumber\\
&=&-B_1 m_0 M \bar{\sigma}^d \chi_0 \Delta_0 \int d\mathbf{V}_1 \int d\mathbf{V}_2f_0^{(0)}(\mathbf{V}_1)f^{(0)}(\mathbf{V}_2)
g_{12}^{-1}\left(\mathbf{g}_{12}\cdot \mathbf{V}_2\right).
\eeqa
To evaluate the integral \eqref{b13}, we approximate $f_0^{(0)}(\mathbf{V}_1)$ and $f^{(0)}(\mathbf{V}_2)$ by their Maxwellian forms:
\beq
\label{b13.1}
f_0^{(0)}(\mathbf{V}_1)\to n_0 \left(\frac{m_0}{2\pi T_0^{(0)}}\right)^{d/2} \exp \left(-\frac{m_0 V_1^2}{2T_0^{(0)}}\right), \quad 
f^{(0)}(\mathbf{V}_2)\to n \left(\frac{m}{2\pi T}\right)^{d/2} \exp \left(-\frac{m V_2^2}{2T}\right).
\eeq
Using these Maxwellian distributions, Eq.\ \eqref{b13} can be expressed as
\beq
\label{b14}
I_{D_T}^{(1)}=-B_1 \pi^{-d} m_0 M n\bar{\sigma}^d n_0 \chi_0 \Delta_0^* v_0^2 \theta^{d/2} I_{D_T}^{(1)*},
\eeq
where the (dimensionless) integral $I_{D_T}^{(1)*}$ is
\beq
\label{b15}
I_{D_T}^{(1)*}=\int d\mathbf{c}_1 \int d\mathbf{c}_2\; g_{12}^{*-1}\left(\mathbf{g}_{12}^*\cdot \mathbf{c}_2\right)e^{-\theta c_1^2-c_2^2}.
\eeq
Here, $\mathbf{g}_{12}^*=\mathbf{g}_{12}/v_0$, $\mathbf{c}_1=\mathbf{V}_1/v_0$,  $\mathbf{c}_2=\mathbf{V}_2/v_0$, and $\theta=m_0 T/m T_0^{(0)}$. The integral \eqref{b15} can be performed by the change of variables $\mathbf{x}=\mathbf{c}_1-\mathbf{c}_2$ and $\mathbf{y}=\theta \mathbf{c}_1+\mathbf{c}_2$, with the Jacobian $(1+\theta)^{-d}$. The result is
\beq
\label{b16}
I_{D_T}^{(1)*}=-\pi^d \frac{\Gamma\left(\frac{d+1}{2}\right)}{\Gamma\left(\frac{d}{2}\right)}
\theta^{\frac{1-d}{2}}(1+\theta)^{-1/2}.
\eeq
With this result, the final expression of $I_{D_T}^{(1)}$ is
\beq
\label{b17}
I_{D_T}^{(1)}=\frac{2^d d}{\sqrt{\pi}}M_0 \left(\frac{\bar{\sigma}}{\sigma}\right)^{d}\phi \chi_0 \left(\frac{\theta}{1+\theta}\right)^{1/2}\Delta_0^*n_0 T,
\eeq
where Eq.\ \eqref{1.14} has been employed. According to Eqs.\ \eqref{b11} and \eqref{b17}, the expression of $I_{D_T}$ can be finally written as
\beq
\label{b18}
I_{D_T}=2^{d}d \left(\frac{\bar{\sigma}}{\sigma}\right)^{d} \phi \chi_0 M_0 n_0 T \left[\frac{1+\al_0}{2}+\frac{\Delta_0^*}{\sqrt{\pi}}
 \left(\frac{\theta}{1+\theta}\right)^{1/2}\right].
\eeq

We consider now the collisional integral
\beq
\label{b19}
I_{D}\equiv \int d\mathbf{v}\; m_0 \mathbf{V}\cdot \boldsymbol{\mathcal{K}}_0\Big[f^{(0)}\Big].
\eeq
By employing the property \eqref{b2}, one easily gets
\beqa
\label{b20}
I_{D}&=&
m_0M\bar{\sigma}^d \chi_0\int d\mathbf{V}_1 \int d\mathbf{V}_2\int d\widehat{\boldsymbol{\sigma}}
\Theta (\widehat{{\boldsymbol {\sigma}}}\cdot {\bf g}_{12})
(\widehat{\boldsymbol {\sigma}}\cdot {\bf g}_{12})f_0^{(0)}(\mathbf{V}_1)f^{(0)}(\mathbf{V}_2)
\Big[\left(1+\alpha_{0}\right)(\widehat{{\boldsymbol {\sigma }}}\cdot \mathbf{g}_{12})+2\Delta_{0}\Big]
\nonumber\\
&=&m_0M\bar{\sigma}^d \chi_0\int d\mathbf{V}_1 \int d\mathbf{V}_2 f_0^{(0)}(\mathbf{V}_1)f^{(0)}(\mathbf{V}_2)\Big[B_2g_{12}^2\left(1+\alpha_{0}\right)+2 B_1 g_{12} \Delta_0\Big].
\eeqa
As in the case of the integral $I_{D_T}$, $I_D$ can be split in two parts; one of them has been already computed when $\Delta_0=0$. Thus,
\beq
\label{b21}
I_{D}=I_{D}^{(0)}+I_{D}^{(1)},
\eeq
where \cite{GHD07}
\beq
\label{b22}
I_{D}^{(0)}=2^{d-1}d \left(\frac{\bar{\sigma}}{\sigma}\right)^{d}\phi \chi_0 M_0 \left(\frac{1+\theta}{\theta}\right)\left(1+\al_0\right)n_0 T,
\eeq
and
\beq
\label{b23}
I_{D}^{(1)}=\frac{2 \pi^{(d-1)/2}}{\Gamma\left(\frac{d+1}{2}\right)}
m_0 M \bar{\sigma}^d \chi_0 \Delta_0 \int d\mathbf{V}_1 \int d\mathbf{V}_2\; g_{12} f_0^{(0)}(\mathbf{V}_1)f^{(0)}(\mathbf{V}_2).
\eeq
The integral $I_{D}^{(1)}$ can be also evaluated by replacing $f_0^{(0)}(\mathbf{V}_1)$ and $f^{(0)}(\mathbf{V}_2)$ by their Maxwellian approximations \eqref{b13.1}. The result is   
\beq
\label{b24}
I_{D}^{(1)}=\frac{2^{d+1} d}{\sqrt{\pi}} \left(\frac{\bar{\sigma}}{\sigma}\right)^{d} \phi \chi_0 M_0 \Delta_0^*\left(\frac{1+\theta}{\theta}\right)^{1/2}n_0T.
\eeq
With this result, the expression of $I_{D}$ can be written as 
\beq
\label{b25}
I_{D}=2^d d \left(\frac{\bar{\sigma}}{\sigma}\right)^{d} \phi \chi_0 M_0 \left(\frac{1+\theta}{\theta}\right)n_0 T\Bigg[\frac{1+\al_0}{2}+\frac{2\Delta_0^*}{\sqrt{\pi}}
\left(\frac{\theta}{1+\theta}\right)^{1/2}\Bigg].
\eeq
\end{widetext}

\bibliography{diffusionDelta}

\end{document}